\documentclass{article}
\usepackage{amsfonts}

\usepackage{amsmath,amssymb}
\usepackage{graphicx}
\usepackage{color}

\newtheorem{fact}{Fact}

\newtheorem{theorem}{Theorem}
\newtheorem{lemma}{Lemma}

\newtheorem{definition}{Definition}

\newtheorem{remark}{Remark}

\newtheorem{example}{Example}

\newcommand{\Z}{\ensuremath{\mathbb Z}}

\newcommand{\done}{\hfill $\Box$ }

\newsavebox{\tablebox}

\newcommand{\ls}[1]
    {\dimen0=\fontdimen6\the\font\lineskip=#1\dimen0
     \advance\lineskip.5\fontdimen5\the\font
     \advance\lineskip-\dimen0
     \lineskiplimit=0.9\lineskip
     \baselineskip=\lineskip
     \advance\baselineskip\dimen0
     \normallineskip\lineskip\normallineskiplimit\lineskiplimit
     \normalbaselineskip\baselineskip
     \ignorespaces}

\begin{document}

\bibliographystyle{abbrv}

\title{Large Zero Autocorrelation Zone of Golay Sequences and $4^q$-QAM
Golay Complementary Sequences }
\author{Guang Gong$^1$  Fei Huo$^1$ and Yang Yang$^{2,3}$  \\
$^1$Department of Electrical and Computer Engineering
University of Waterloo \\
Waterloo, Ontario N2L 3G1, CANADA \\
$^2$Institute of Mobile Communications, Southwest
Jiaotong University \\
Chengdu, 610031, P.R. CHINA \\
Email: ggong@calliope.uwaterloo.ca, fhuo@engmail.uwaterloo.ca,\\
 yang$\_$data@yahoo.cn  \\
}

\date{}
 \maketitle

\thispagestyle{plain} \setcounter{page}{1}

\begin{abstract}

Sequences with good correlation properties have been widely adopted
in modern communications, radar and sonar applications. In this
paper, we present our new findings on some constructions of single
$H$-ary Golay sequence and $4^q$-QAM Golay complementary sequence
with a large zero autocorrelation zone, where $H\ge 2$ is an
arbitrary even integer and $q\ge 2$ is an arbitrary integer. Those
new results on Golay sequences and QAM Golay complementary sequences
can be explored during synchronization and detection at the receiver
end and thus improve the performance of the communication system.


\end{abstract}

\ls{1.5}
\section{Introduction}

\footnotetext[3]{Yang Yang is current a visiting Ph. D student (Oct.
2010- Sep. 2012) in the ECE, University of Waterloo.}

In modern communications, sequences with good correlation properties
are desired for receiver synchronization and detection purposes. In
1961, Golay proposed the idea of aperiodic complementary sequence
pairs \cite{Golay}, of which the sum of out-of-phase aperiodic
autocorrelation equals to zero. Later on, Davis and Jedwab
formulated a method for constructing Golay complementary pairs by
using quadratic generalized boolean functions \cite{DJ99}. Due to
this correlation property, Golay sequences have been proposed to
construct Hadamard matrix for direct sequence code division multiple
access (DS-CDMA) system \cite{seberry}, and to control the peak
envelope power (PEP) in orthogonal frequency-division multiplexing
(OFDM) system \cite{van Nee,Wilkinson95,Wilkinson96,Wulich}.

The utilization of Golay sequences in the two above scenarios are
based on the property that the sum of out-of-phase autocorrelation
of the pair equals to zero. However, synchronization and detection
of the signal is equivalent to computing its own autocorrelation. In
this case, investigation of the autocorrelation of single sequence
is of our interest in this paper. This is also the case with
conventional CDMA and quasi-synchronous code-division
multiple-access (QS-CDMA) system.

QS-CDMA differs from conventional CDMA system \cite{GG05} in that it
allows a small time delay in the arrival signals of different users.
In this case, sequences with low or zero correlations centered at
the origin are desired to eliminate or reduce the multiple access
and multipath interference at the receiver end during detection.
Such sequences are called low correlation zone (LCZ) and zero
correlation zone (ZCZ) sequences respectively \cite{Long}. As a
result, the construction of new LCZ or ZCZ sequences for QS-CDMA
system has received researchers's much  attention
\cite{DF00,FH00,Hayashi02,Hayashi04,HG10,RC06,RC08,Tang00,Tang06,ZTG10}.

Our motivation is to examine the correlation properties of Golay
sequences and quadrature amplitude modulation (QAM) Golay
complementary sequences when it is being utilized for signal
detection and synchronization purposes in applications such as CDMA
and conventional linear time invariant (LTI) system. More
specifically, if single Golay sequence or QAM Golay complementary
sequence inherits some fixed or attractive autocorrelation property
which can be exploited during detection and thus improves the
performance of the system. Please refer to
\cite{CVT,Lee,Li2008,CLH,Li2010} more details on QAM Golay
complementary sequences. In this paper, we will present our findings
on several constructions of Golay sequences and QAM Golay
complementary sequences with a zero autocorrelation zone (ZACZ) of
length approximate an half, a quarter or one eighth of their
periods.

This paper is organized as follows. In Section $2$, we provide the
necessary preliminary materials required in the later sections. In
Sections $3$ and $4$, we show the large ZACZ of Golay sequences and
QAM Golay complementary sequences. In Section $5$, we demonstrate
the ZACZ with concrete examples. Finally, we conclude our paper in
Section $6$.

\section{Definitions and Preliminaries }

Let $H\ge 2$ be an arbitrary integer and $\xi$ be the primitive
$H$-th root of unity, i.e., $\xi=\exp(2\pi\sqrt{-1}/H)$. For a
sequence $a=(a_0,a_1,\cdots,a_{N-1})$ over $Z_H$ with period $N$,
its  {\em aperiodic autocorrelation function} and {\em periodic
autocorrelation function} are respectively defined by
\[
C_{a}(\tau)=\sum_{i=0}^{N-1-\tau} \xi^{a_i-a_{i+\tau}}, \tau=0, 1,
\cdots,
\]
and
\[
R_{a}(\tau)=\sum_{i=0}^{N-1} \xi^{a_i-a_{i+\tau}}, \tau=0, 1,
\cdots.
\]

\begin{definition}Let $\delta_1$ and $\delta_2$ be two integers with
$0<\delta_1<\delta_2<N$ and denote $L=\delta_2-\delta_1+1$. If the
periodic autocorrelation function of $a$ is equal to zero with a
range $\delta_1\le \tau\le \delta_2$, then the sequence $a$ has a
{\it zero autocorrelation zone} (ZACZ) of length $L$.
\end{definition}

  This definition is a
variation of the definition given in \cite{DF00}.

Let $a$ and $b$ be two sequences over $Z_H$ with period $N$. The
sequences $a$ and $b$ are called a {\em Golay complementary pair} if
$C_{a}(\tau)+C_{b}(\tau)=0$ for any $1\le \tau\le N-1$. Any one of
them is called a {\em Golay sequence}.

A {\em generalized Boolean function} $f(x_1,\cdots,x_m)$ with $m$
variables is a mapping from $\{0,1\}^m$ to $\Z_H$, which has a
unique representation as a multiple polynomial over $\Z_H$ of the
special form:
$$f(x_1,\cdots,x_m)=\sum_{I\in \{1,\cdots,m\}}a_I\prod_{i\in I}x_i,\,\,a_I\in \Z_H, x_i\in \{0,1\}.$$
This is called the {\em algebraic normal form} of $f$. The {\em
algebraic degree} is defined by the maximum value of the size of the
set $I$ with $a_I\ne 0$.

Let $(i_1,\cdots,i_m)$ be the binary representation of the integer
$i=\sum_{k=1}^mi_k2^{m-k}$. The {\em truth table} of a Boolean
function $f(x_1,\cdots,x_m)$ is a binary string of length $2^m$,
where the $i$-th element of the string is equal to
$f(i_1,\cdots,i_m)$. For example, $m=3$, we have
\begin{eqnarray*}f&=&(f(0,0,0),f(0,0,1),f(0,1,0),f(0,1,1),f(1,0,0),f(1,0,1),f(1,1,0),f(1,1,1)).\end{eqnarray*}

In the following, we introduce some notations. We always assume that
$m\ge 4$ is an integer and $\pi$ is a permutation from
$\{1,\cdots,m\}$ to itself.

\begin{definition}\label{de golay}
Define a sequence $a=\{a_i\}_{i=0}^{2^m-1}$ over $\Z_H$, whose
elements are given by
\begin{eqnarray}\label{eq a}a_i={H\over 2}\sum_{k=1}^{m-1}i_{\pi(k)}i_{\pi(k+1)}+\sum_{k=1}^m
c_ki_k+c_0,\end{eqnarray} where $c_i\in \Z_H$, $i=0,1,\cdots,m$.
\end{definition}

When $H=2^h$, $h\ge 1$ an integer, Davis and Jedwab proved that
$\{a_i\}$ and $\{a_i+2^{h-1}i_{\pi(1)}+c'\}$ form a Golay
complementary pair for any $c'\in \Z_{2^h}$ in the Theorem $3$ of
\cite{DJ99}. Later on, Paterson generalized this result by replacing
$\Z_{2^h}$ with $\Z_H$ \cite{Paterson}, where $H\ge 2$ is an
arbitrary even integer.

\begin{fact}[Corollary 11, \cite{Paterson}]\label{th Davis}
Let $a=\{a_i\}_{i=0}^{2^m-1}$ be the sequence given in Definition
\ref{de golay}. Then the pair of the sequences $a_i$ and
$a_i+{H\over 2}i_{\pi(1)}+c'$ form a Golay complementary pair for
any $c'\in \Z_H$.
\end{fact}

We define
\begin{eqnarray*}
a_{i,0}&=&2\sum_{k=1}^{m-1}i_{\pi(k)}i_{\pi(k+1)}+\sum_{k=1}^mc_ki_k+c_0\\
b_{i,0}&=&a_{i,0}+\mu_i\\
a_{i,e}&=&a_{i,0}+s_{i,e}\\
b_{i,e}&=&b_{i,0}+s_{i,e}=a_{i,e}+\mu_i,1\le e\le q-1,
\end{eqnarray*}
where $c_k\in \Z_4$, $k=0,1,\cdots,m$, and $s_{i,e}$ and $\mu_i$ are
defined as one of the following cases:
\begin{enumerate}
\item [1.]
$s_{i,e}=d_{e,0}+d_{e,1}i_{\pi(m)}$, $\mu_i=2i_{\pi(1)}$ for any
$d_{e,0},d_{e,1}\in \Z_4$.
\item [2.]
$s_{i,e}=d_{e,0}+d_{e,1}i_{\pi(1)}$, $\mu_i=2i_{\pi(m)}$ for any
$d_{e,0},d_{e,1}\in \Z_4$.
\item [3.]
$s_{i,e}=d_{e,0}+d_{e,1}i_{\pi(w)}+d_{e,2}i_{\pi(w+1)}$,
$2d_{e,0}+d_{e,1}+d_{e,2}=0$, $\mu_i=2i_{\pi(1)}$, or
$\mu_i=2i_{\pi(m)}$ for any $d_{e,0},d_{e,1},d_{e,2}\in \Z_4$ and
$1\le w\le m-1$.
\end{enumerate}
We construct a pair of $4^q$-QAM sequences $A=\{A_i\}_{i=0}^{2^m-1}$
and $B=\{B_i\}_{i=0}^{2^m-1}$ as follows:
\begin{align}\label{eq qam}
\begin{array}{rcl}
A_i&=&\gamma\sum\limits_{e=0}^{q-1}r_j\xi^{a_{i,e}}\\
B_i&=&\gamma\sum\limits_{e=0}^{q-1}r_j\xi^{b_{i,e}},
\end{array}
\end{align}
where $\gamma=e^{j\pi/4}$, $\xi=\sqrt{-1}$, and $r_p={2^{q-1-j}\over
\sqrt{(4^q-1)/3}}$, $ a_{i,e},b_{i,e}\in \Z_4$, $0\le e\le q-1$.

\begin{fact}[Theorem 2, \cite{Li2010}]\label{th Li}
The two sequence $A$ and $B$ form a $4^q$-QAM Golay complementary
pair. Furthermore, for $q=2$, $A$ and $B$ become $16$-QAM Golay
complementary pair which are constructed by Chong, Venkataramani and
Tarokh in \cite{CVT}; For $q=3$, $A$ and $B$ become $64$-QAM Golay
complementary pair which are presented by Lee and Golomb in
\cite{Lee}.
\end{fact}

\begin{remark}
Note that there are some typos and missing cases in the original
publication of \cite{CVT} and \cite{Lee}. However, those are
corrected in \cite{Li2008}. Some additional cases about $64$-QAM
Golay complementary sequences, which are not of those forms above,
are presented in \cite{CLH}.
\end{remark}

In the remaining of this paper, we adopt the following notations:
For an integer $\tau$, $1\le \tau\le 2^m-1$, two integers $i$ and
$i'$, $0\le i,i',j,j'<2^m$, we set $j=(i+\tau)~\bmod~2^m$ and
$j'=(i'+\tau)~\bmod~2^m$, and let $(i_1,\cdots,i_m)$,
$(i'_1,\cdots,i'_m)$, $(j_1,\cdots,j_m)$ and $(j'_1,\cdots,j'_m)$ be
the binary representations of $i$, $i'$, $j$, $j'$, respectively.

\section{Zero Autocorrelation Zone of Golay Sequences}

In this section, we will study the ZACZ of Golay sequences.

\subsection{Pre-described Conditions}

In this subsection, we list $3$ sets of conditions on permutations
$\pi$ and affine transformation $\sum_{k=1}^mc_ki_k+c_0$.

\begin{enumerate}
\item
[(A)] \begin{enumerate}
\item [(1)] $\pi(1)=1$, $\pi(2)=2$ and $2c_1=0$.
\item [(2)] $\pi(2)=2$, $\pi(3)=1$, $\pi(4)=3$,
$2c_1=0$ and $c_1=2c_2$.
\item [(3)] $\pi(1)=2$, $\pi(2)=1$, $\pi(3)=3$, $2c_1=0$ and $c_1=2c_2+t$, where
$$t=\left\{\begin{array}{ll}
{H\over 2}, & \mbox{ for Golay sequences defined by equality \eqref{eq a}}\\
2, & \mbox{ for QAM Golay complementary sequences defined by
equality \eqref{eq qam}}.\end{array}\right.$$
\end{enumerate}
\item [(B)]
$\pi(1)=2$, $\pi(2)=1$, $\pi(3)=3$, $2c_1=0 $ and $c_1=2c_2$.
\item [(C)] \begin{enumerate}
\item [(1)] $\pi(1)=1$, $\pi(2)=3$, $\pi(3)=2$ and $2c_1=0$.
\item [(2)] $\pi(1)=1$, $\pi(2)=3$, $\pi(m)=2$ and $2c_1=0$.
\item [(3)] $\pi(1)=2$, $\pi(2)=4$, $\pi(3)=1$, $\pi(4)=3$,
$2c_1=0$ and $c_1=2c_2$.
\item [(4)] $\pi(1)=2$, $\pi(2)=3$, $\pi(3)=1$, $\pi(4)=4$, $2c_1=0$ and $c_1=2c_2$.
\end{enumerate}
\end{enumerate}

Define a mapping $\pi'(k)=\pi(m+1-k)$, $k\in \{1,\cdots,m\}$.
Replacing $\pi$ by $\pi'$, the above three sets of the conditions on
permutations $\pi$ and affine transformation
$\sum_{k=1}^mc_ki_k+c_0$ above can be written as follows.

\begin{enumerate}
\item
[(A')] \begin{enumerate}
\item [(1)] $\pi(m)=1$, $\pi(m-1)=2$ and $2c_1=0$.
\item [(2)] $\pi(m-1)=2$, $\pi(m-2)=1$, $\pi(m-3)=3$,
$2c_1=0$ and $c_1=2c_2$.
\item [(3)] $\pi(m)=2$, $\pi(m-1)=1$, $\pi(m-2)=3$, $2c_1=0$ and $c_1=2c_2+t$, where
$$t=\left\{\begin{array}{ll}
{H\over 2}, & \mbox{ for Golay sequences defined by equality \eqref{eq a}}\\
2, & \mbox{ for QAM Golay complementary sequences defined by
equality \eqref{eq qam}}.\end{array}\right.$$
\end{enumerate}
\item [(B')]
$\pi(m)=2$, $\pi(m-1)=1$, $\pi(m-2)=3$, $2c_1=0 $ and $c_1=2c_2$.
\item [(C')] \begin{enumerate}
\item [(1)] $\pi(m)=1$, $\pi(m-1)=3$, $\pi(m-1)=2$ and $2c_1=0$.
\item [(2)] $\pi(m)=1$, $\pi(m-1)=3$, $\pi(1)=2$ and $2c_1=0$.
\item [(3)] $\pi(m)=2$, $\pi(m-1)=4$, $\pi(m-2)=1$, $\pi(m-3)=3$,
$2c_1=0$ and $c_1=2c_2$.
\item [(4)] $\pi(m)=2$, $\pi(m-1)=3$, $\pi(m-2)=1$, $\pi(m-3)=4$, $2c_1=0$ and $c_1=2c_2$.
\end{enumerate}
\end{enumerate}

\subsection{Main Results}

\begin{figure}[ht]
\begin{minipage}[b]{1.0\linewidth}
\centering
\includegraphics[scale=1]{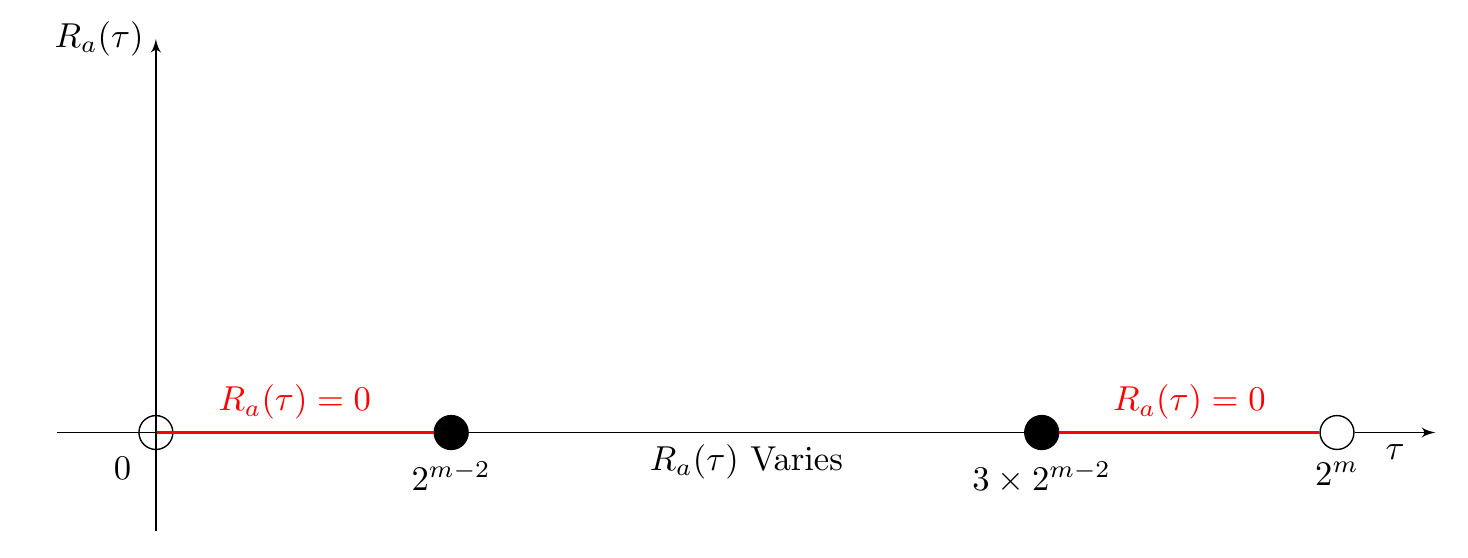}
\caption{The Zero Autocorrelation Zone of Golay Sequence $a$ Defined
by \eqref{eq a} and Condition (A)} \label{picture 1}
\end{minipage}
\end{figure}

\begin{theorem}\label{th 1}
If the Golay sequence $a$, defined by Definition \ref{de golay},
satisfies one of the condition listed in (A) or (A'), then the
sequence $a$ has the following property:
$$R_a(\tau)=0,\,\,\,\,\,\,\, \tau\in (0,2^{m-2}]\cup [3\cdot
2^{m-2},2^m).$$ In other words, in one period $[0,2^m)$, it has two
zero autocorrelation zones of length $2^{m-2}$, given by
$(0,2^{m-2}]$ and $[3\cdot 2^{m-2},2^m)$, shown in Figure
\ref{picture 1}.
\end{theorem}

\begin{figure}[ht]
\begin{minipage}[b]{1.0\linewidth}
\centering
\includegraphics[scale=1]{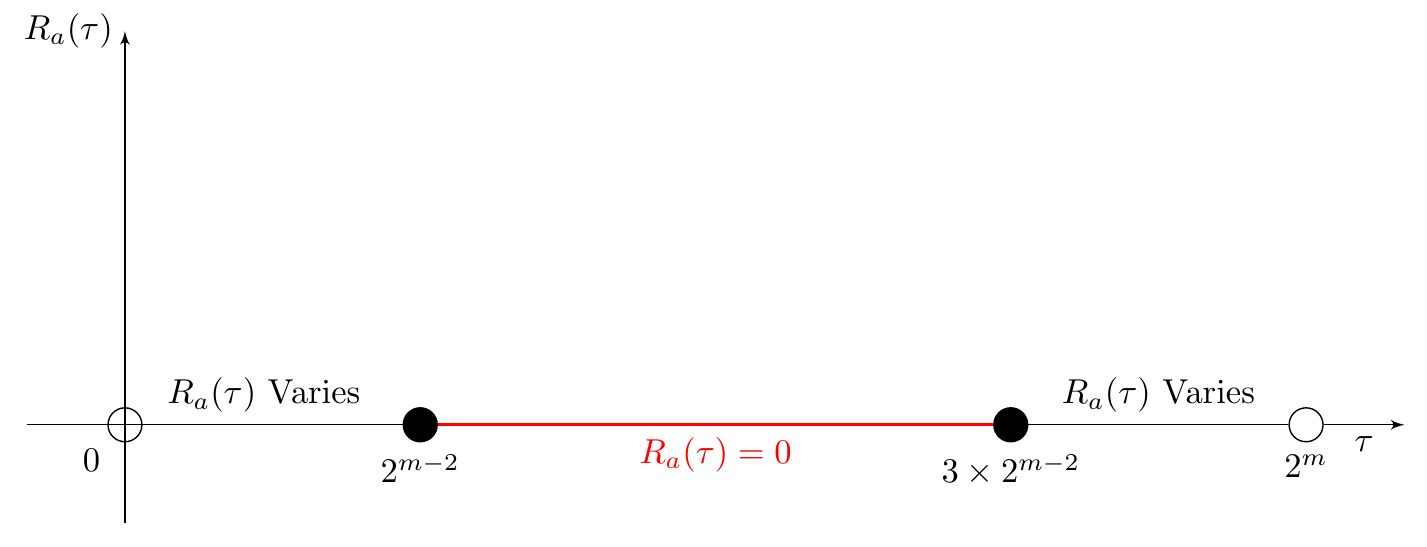}
\caption{The Zero Autocorrelation Zone of Golay Sequence $a$ Defined
by \eqref{eq a} and Condition (B)} \label{picture 2}
\end{minipage}
\end{figure}

\begin{theorem}\label{th 2}
If the Golay sequence $a$, defined by Definition \ref{de golay},
satisfies one of the condition listed in (B) or (B'), then the
sequence $a$ has the following property: $$R_a(\tau)=0,
\,\,\,\,\,\,\, \tau\in [2^{m-2},3\cdot 2^{m-2}].$$ In other words,
in one period $[0,2^m)$, it has a zero autocorrelation zone of
length $2^{m-1}+1$, given by $[2^{m-2},3\cdot 2^{m-2}]$, shown in
Figure \ref{picture 2}.
\end{theorem}

\begin{figure}[ht]
\begin{minipage}[b]{1.0\linewidth}
\centering
\includegraphics[scale=1]{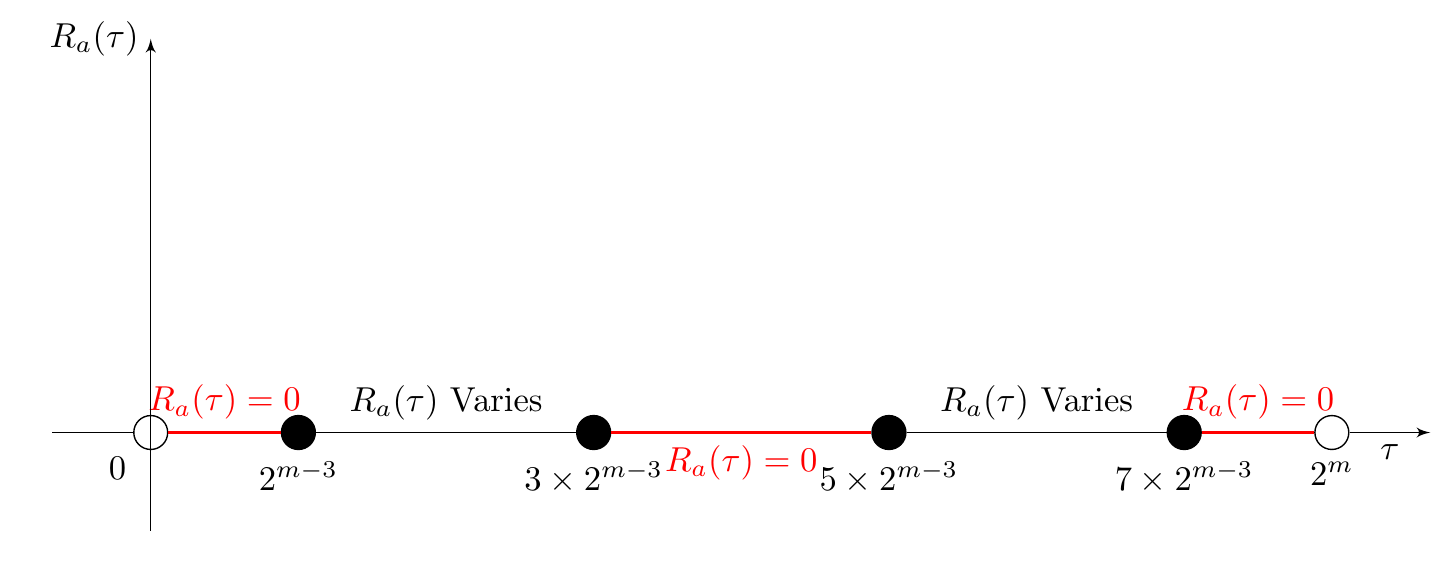}
\caption{The Zero Autocorrelation Zone of Golay Sequence $a$ Defined
by \eqref{eq a} and Condition (C)} \label{picture 3}
\end{minipage}
\end{figure}

\begin{theorem}\label{th 3}
If the Golay sequence $a$, defined by Definition \ref{de golay},
satisfies one of the condition listed in (C) or (C'), then the
sequence $a$ has the following property: $$R_a(\tau)=0,
\,\,\,\,\,\tau\in (0,2^{m-3}]\cup [3\cdot 2^{m-3}, 5\cdot
3^{m-3}]\cup [7\cdot 2^{m-3},2^m).$$ In other words, in one period
$[0,2^m)$, it has three zero autocorrelation zones of respective
length $2^{m-3}$, $2^{m-2}+1$, $2^{m-3}$, given by $(0,2^{m-3}]$,
$[3\cdot 2^{m-3}, 5\cdot 3^{m-3}]$ and $[7\cdot 2^{m-3},2^m)$, shown
in Figure \ref{picture 3}.
\end{theorem}

\subsection{Proofs of the Main Results}

The set $\{i: 0\le i\le 2^m-1\}$ can be divided into the following
three disjoint subsets:
\begin{eqnarray*}
I_1(\tau)&=&\{0\le i\le 2^m-1:i_{\pi(1)}=j_{\pi(1)}\};\\
I_2(\tau)&=&\{0\le i\le 2^m-1:i_{\pi(1)}\ne j_{\pi(1)},i_{\pi(m)}=j_{\pi(m)}\};\\
I_3(\tau)&=&\{0\le i\le 2^m-1: i_{\pi(1)}\ne
j_{\pi(1)},i_{\pi(m)}\ne j_{\pi(m)}\}.
\end{eqnarray*}
Then the periodic autocorrelation function $R_a(\tau)$ can be
written as
\begin{eqnarray}\label{eq autocorrelation}
R_a(\tau)=\sum\limits_{i=0}^{2^m-1}\xi^{a_i-a_j}=\sum\limits_{i\in
I_1(\tau)}\xi^{a_i-a_j}+\sum\limits_{i\in I_2(\tau)}\xi^{a_i-a_j}
+\sum\limits_{i\in I_3(\tau)}\xi^{a_i-a_j}.\end{eqnarray}

\begin{lemma}\label{le 1}
For any Golay sequence $a$ given by Definition \ref{de golay}, for
an integer $\tau$, $1\le \tau\le 2^m-1$, we have
\begin{eqnarray*}
\sum\limits_{i\in I_1(\tau)}\xi^{a_i-a_j}&=&0.\end{eqnarray*}
\end{lemma}

{\em Proof:}  Since $j=(i+\tau)~\bmod~2^m\ne i$, for each $i\in
I_1(\tau)$, we can define $v$ as follows:
$$v=\min\{1\le k\le m: i_{\pi(k)}\ne j_{\pi(k)}\}.$$
From the definition of $I_1(\tau)$, it is immediately seen that
$v\ge 2$. Let $i'$ and $j'$ be two integers with binary
representations defined by
\begin{eqnarray*}i'_{\pi(k)}=\left\{\begin{array}{ll} i_{\pi(k)}, &  k\ne v-1\\
1-i_{\pi(k)}, &  k=v-1\end{array}\right.\end{eqnarray*} and
\begin{eqnarray*}j'_{\pi(k)}=\left\{\begin{array}{ll} j_{\pi(k)}, &  k\ne v-1\\
1-j_{\pi(k)}, &  k=v-1.\end{array}\right.\end{eqnarray*} In other
words, $i'$ and $j'$ are obtained from $i$ and $j$ by ``flipping''
the $(v-1)$-th bit in $(i_{\pi(1)},\cdots,i_{\pi(m)})$ and
$(j_{\pi(1)},\cdots,j_{\pi(m)})$. We can derive the following
results.
\begin{enumerate}
\item [1)] $j'-i'=j-i\equiv \tau~\bmod~2^m$ for any $i\in I_1(\tau)$.
\item [2)] $i'_{\pi(1)}=j'_{\pi(1)}$.
\item [3)] The mapping $i\rightarrow i'$ is a one-to-one
mapping.
\end{enumerate}
Hence $i'$ enumerates $I_1(\tau)$ as $i$ ranges over $I_1(\tau)$.
For any given $i\in I_1(\tau)$, we have
\begin{eqnarray*}
a_i-a_j-a_{i'}+a_{j'}={H\over 2}.
\end{eqnarray*}
This equality implies $\xi^{a_i-a_j}/\xi^{a_{i'}-a_{j'}}=-1$, thus
$$\xi^{a_i-a_j}+\xi^{a_{i'}-a_{j'}}=0.$$

Hence we have
$$2\sum\limits_{i\in
I_1(\tau)}\xi^{a_i-a_j}=\sum\limits_{i\in
I_1(\tau)}\xi^{a_i-a_j}+\sum\limits_{i'\in
I_1(\tau)}\xi^{a_{i'}-a_{j'}}=\sum\limits_{i\in
I_1(\tau)}\left(\xi^{a_i-a_j}+\xi^{a_{i'}-a_{j'}}\right)=0.$$ Thus
it follows that $\sum_{i\in I_1(\tau)}\xi^{a_i-a_j}=0$.

\done

\begin{lemma}\label{le 2}
For any Golay sequence $a$ given by Definition \ref{de golay}, for
an integer $\tau$, $1\le \tau\le 2^m-1$, we have
\begin{eqnarray*}
\sum\limits_{i\in I_2(\tau)}\xi^{a_i-a_j}&=&0.\end{eqnarray*}
\end{lemma}

{\em Proof:} For any $i\in I_2(\tau)$, let $i'$ and $j'$ be the two
integers with binary representations defined by
\begin{eqnarray*}i'_{\pi(k)}=1-j_{\pi(k)},\,\,\  k=1,\cdots,m\end{eqnarray*} and
\begin{eqnarray*}j'_{\pi(k)}=1-i_{\pi(k)},\,\, \  k=1,\cdots,m.\end{eqnarray*}
We have the following results:
\begin{enumerate}
\item [1)] $j'-i'= j-i\equiv \tau~\bmod~2^m$ for any $i\in
I_2(\tau)$.
\item [2)] $i'_{\pi(1)}\ne j'_{\pi(1)}$ and $i'_{\pi(m)}=
j'_{\pi(m)}$.
\item [3)] The mapping $i\rightarrow i'$ is a one-to-one
mapping. This together with the two facts above implies that $i'$
enumerates $I_2(\tau)$ as $i$ ranges over $I_2(\tau)$.

\item [4)] $a_i-a_j-a_{i'}+a_{j'}={H\over 2}$. This
implies $\xi^{a_i-a_j}/\xi^{a_{i'}-a_{j'}}=-1$, and then
$\xi^{a_i-a_j}+\xi^{a_{i'}-a_{j'}}=0$ for any $i\in I_2(\tau)$.
\end{enumerate}
Hence the conclusion follows immediately.

\done

By Lemmas \ref{le 1} and \ref{le 2}, the periodic autocorrelation
function $R_a(\tau)$ can be reduced as
\begin{eqnarray}\label{eq auto}
R_a(\tau)=\sum\limits_{i\in I_3(\tau)}\xi^{a_i-a_j}.\end{eqnarray}

Now we will present the ZACZ findings of Golay sequences, i.e.,
equality in \eqref{eq auto} is equal to zero.

Note that for the sets $I_1(\tau)$ and $I_2(\tau)$, their proofs are
independent of the choice of permutations $\pi$ and affine
transformations $\sum_{i=1}^mc_ix_i+c_0$. However for the set
$I_3(\tau)$, the proof for each case in Theorem \ref{th 1} is
different.

In order to prove Theorem \ref{th 1}, we need several lemmas on
$\sum_{i\in I_3(\tau)}\xi^{a_i-a_j}=0$ in the three cases.

\begin{lemma}\label{le 3}
Let $a$ be the sequence given by Definition \ref{de golay} and
satisfy the condition (A)-(1). Then we have $i_{\pi(2)}\ne
j_{\pi(2)}$ for any $i\in I_3(\tau)$, i.e.,
\begin{eqnarray*}
I_3(\tau) &=&\{0\le i\le 2^m-1: i_{\pi(1)}\ne
j_{\pi(1)},i_{\pi(2)}\ne j_{\pi(2)},i_{\pi(m)}\ne j_{\pi(m)}\}
\end{eqnarray*}
and $\sum_{i\in I_3(\tau)}\xi^{a_i-a_j}=0$ for any $\tau\in \{k:1\le
k\le 2^{m-2}\}$.
\end{lemma}

{\em Proof: } We can partition $I_3(\tau)$ into the following two
disjoint subsets:
\begin{eqnarray*}
I_4(\tau) &=&\{0\le i\le 2^m-1:i_{\pi(1)}\ne j_{\pi(1)},i_{\pi(2)}=
j_{\pi(2)},i_{\pi(m)}\ne j_{\pi(m)}\};\\
I_5(\tau) &=&\{0\le i\le 2^m-1: i_{\pi(1)}\ne
j_{\pi(1)},i_{\pi(2)}\ne j_{\pi(2)},i_{\pi(m)}\ne j_{\pi(m)}\}.
\end{eqnarray*}

First we will show that $I_4(\tau)$ is an empty set. $i_{\pi(1)}\ne
j_{\pi(1)}$ implies that: (i) $(i_{\pi(1)},j_{\pi(1)})=(0,1)$; or
(ii) $(i_{\pi(1)},j_{\pi(1)})=(1,0)$.

(i) On one hand, note that $j\equiv (i+\tau) \bmod 2^m$ together
with $i<2^{m-1}$ and $\tau\le 2^{m-2}$, we have $j=i+\tau<2^m$. On
the other hand, we have
\begin{eqnarray*}
i+\tau<2^{m-1}+2^{m-2}=j_{\pi(1)}2^{m-1}+j_{\pi(2)}2^{ m-2}\le j
&\Rightarrow& i+\tau<j\end{eqnarray*} which is a contradiction with
$j=i+\tau$.

(ii) Note that $j\equiv (i+\tau) \bmod 2^m$ together with
$j<2^{m-1}<i<2^m$ and $\tau\le 2^{m-2}$, we have $j=i+\tau-2^m$.
Similar as (i), we have
\begin{eqnarray*}
i+\tau&<&(i_{\pi(1)}2^{m-1}+i_{\pi(2)}2^{m-2}+2^{m-2})+2^{m-2}\\
&=&i_{\pi(2)}2^{ m-2}+2^m=j_{\pi(2)}2^{ m-2}+2^m\le
j+2^m\\
&\Rightarrow& i+\tau<j+2^m\end{eqnarray*} which contradicts with
that $j=i+\tau-2^m$.

By the discussion above, we conclude that $I_4(\tau)$ is an empty
set and $I_3(\tau)=I_5(\tau)$. Thus we have
$$\sum_{i\in I_3(\tau)}\xi^{a_i-a_j}=\sum_{i\in I_5(\tau)}\xi^{a_i-a_j}.$$

For the case $i\in I_5(\tau)$, let $i'$ and $j'$ be two integers
with binary representations defined by
\begin{eqnarray*}i'_{\pi(k)}=\left\{\begin{array}{ll} 1-i_{\pi(k)}, & k=1\\
i_{\pi(k)}, & k\ne 1\end{array}\right.\end{eqnarray*} and
\begin{eqnarray*}j'_{\pi(k)}=\left\{\begin{array}{ll} 1-j_{\pi(k)}, & k=1\\
j_{\pi(k)}, & k\ne 1.\end{array}\right.\end{eqnarray*} We have the
following assertions:
\begin{enumerate}
\item [1)] $j'-i'\equiv j-i\equiv \tau~\bmod~2^m$ for any $i\in
I_5(\tau)$.
\item [2)] $i'$
satisfies $i'_{\pi(1)}\ne j'_{\pi(1)}$, $i'_{\pi(2)}\ne
j'_{\pi(2)}$, and $i'_{\pi(m)}\ne j'_{\pi(m)}$, i.e., $i'\in
I_5(\tau)$.
\item [3)] The mapping $i\rightarrow i'$ is a one-to-one
mapping. This together with the two facts above indicates that $i'$
enumerates $I_5(\tau)$ as $i$ ranges over $I_5(\tau)$.

\item [4)]
$a_i-a_j-(a_{i'}-a_{j'})={H\over
2}(i_{\pi(2)}+j_{\pi(2)})+2c_1(i_{\pi(1)}-j_{\pi(1)})={H\over 2}$.
This implies $\xi^{a_i-a_j}+\xi^{a_{i'}-a_{j'}}=0$ for any  $i\in
I_5(\tau)$.
\end{enumerate}
Hence we have $\sum_{i\in I_3(\tau)}\xi^{a_i-a_j}=\sum_{i\in
I_5(\tau)}\xi^{a_i-a_j}=0$.

\done

\begin{lemma}\label{le 4}
Let $a$ be the sequence given by Definition \ref{de golay} and
satisfy the condition (A)-(2). Then we have $\sum_{i\in
I_3(\tau)}\xi^{a_i-a_j}=0$ for any $\tau\in \{k:1\le k\le
2^{m-2}\}$.
\end{lemma}

{\em Proof: } We partition $I_3(\tau)$ into the following four
disjoint subsets:
\begin{eqnarray*}
I_6(\tau)&=&\{0\le i\le 2^m-1: i_{\pi(1)}\ne j_{\pi(1)}, i_{\pi(2)}=
j_{\pi(2)}, i_{\pi(3)}\ne j_{\pi(3)}, i_{\pi(4)}= j_{\pi(4)},
i_{\pi(m)}\ne j_{\pi(m)}\},\\
I_7(\tau)&=&\{0\le i\le 2^m-1: i_{\pi(1)}\ne j_{\pi(1)}, i_{\pi(2)}=
j_{\pi(2)}, i_{\pi(3)}= j_{\pi(3)}, i_{\pi(4)}= j_{\pi(4)},
i_{\pi(m)}\ne j_{\pi(m)}\},\\
 I_8(\tau)&=&\{0\le i\le 2^m-1:
i_{\pi(1)}\ne j_{\pi(1)}, i_{\pi(2)}+j_{\pi(2)}+i_{\pi(4)}+
j_{\pi(4)}=1, i_{\pi(m)}\ne
j_{\pi(m)}\}, \mbox{ and }\\
I_9(\tau)&=&\{0\le i\le 2^m-1: i_{\pi(1)}\ne j_{\pi(1)},
i_{\pi(2)}\ne j_{\pi(2)}, i_{\pi(4)}\ne j_{\pi(4)}, i_{\pi(m)}\ne
j_{\pi(m)}\}.
\end{eqnarray*}

Similar to the analysis of $I_4(\tau)$, we have that $I_6(\tau)$ is
an empty set. Thus
\begin{eqnarray}\label{eq 2}\sum_{i\in I_3(\tau)}\xi^{a_i-a_j}=\sum_{i\in I_7(\tau)}\xi^{a_i-a_j}+\sum_{i\in I_8(\tau)}\xi^{a_i-a_j}
+\sum_{i\in I_9(\tau)}\xi^{a_i-a_j}.\end{eqnarray}

Now we will show that
\begin{eqnarray}\label{eq I7}
\sum\limits_{i\in I_7(\tau)}\xi^{a_i-a_j}&=&0\\
\label{eq I8}\sum\limits_{i\in I_8(\tau)}\xi^{a_i-a_j}&=&0\\
\label{eq I9}\sum\limits_{i\in
I_9(\tau)}\xi^{a_i-a_j}&=&0.\end{eqnarray}

For any given $i\in I_7(\tau)$, let $i'$ and $j'$ be two integers
with binary representations defined by
\begin{eqnarray*}i'_{\pi(k)}=\left\{\begin{array}{ll} j_{\pi(k)}, & k=2,3\\
1-j_{\pi(k)}, & \mbox{otherwise} \end{array}\right.\end{eqnarray*}
and
\begin{eqnarray*}j'_{\pi(k)}=\left\{\begin{array}{ll} i_{\pi(k)}, & k=2,3\\
1-i_{\pi(k)}, & \mbox{otherwise.} \end{array}\right.\end{eqnarray*}
We have the following results.
\begin{enumerate}
\item [1)] $j'-i'\equiv j-i\equiv \tau~\bmod~2^m$ for any $i\in
I_7(\tau)$.
\item [2)] $i'$ satisfies $i'_{\pi(1)}\ne j'_{\pi(1)}$, $i'_{\pi(2)}= j'_{\pi(2)}$,
$i'_{\pi(3)}= j'_{\pi(3)}$, $i'_{\pi(4)}= j'_{\pi(4)}$, and
$i'_{\pi(m)}\ne j'_{\pi(m)}$, i.e., $i'\in I_7(\tau)$.
\item [3)] The mapping $i\rightarrow i'$ is a one-to-one
mapping.  This together with the two facts above indicates that $i'$
enumerates $I_7(\tau)$ as $i$ ranges over $I_7(\tau)$.

\item [4)]
$a_i-a_j-(a_{i'}-a_{j'})={H\over 2}(i_{\pi(2)}+j_{\pi(2)}
+i_{\pi(3)}+j_{\pi(3)}+i_{\pi(4)}+j_{\pi(4)}+i_{\pi(m)}+j_{\pi(m)})={H\over
2}$. This implies $\xi^{a_i-a_j}+\xi^{a_{i'}-a_{j'}}=0$ for any
$i\in I_7(\tau)$.
\end{enumerate}
Hence we have equality \eqref{eq I7} holds.

For any given $i\in I_8(\tau)$, let $i'$ and $j'$ be two integers
with binary representations defined by
\begin{eqnarray*}i'_{\pi(k)}=\left\{\begin{array}{ll} j_{\pi(k)}, & k=3\\
1-j_{\pi(k)}, & \mbox{otherwise} \end{array}\right.\end{eqnarray*}
and
\begin{eqnarray*}j'_{\pi(k)}=\left\{\begin{array}{ll} i_{\pi(k)}, & k=3\\
1-i_{\pi(k)}, & \mbox{otherwise.} \end{array}\right.\end{eqnarray*}
We have the following assertions.
\begin{enumerate}
\item [1)] $j'-i'\equiv j-i\equiv \tau~\bmod~2^m$ for any $i\in
I_8(\tau)$.
\item [2)] $i'$
satisfies $i'_{\pi(1)}\ne j'_{\pi(1)}$,
$i'_{\pi(2)}+j'_{\pi(2)}+i'_{\pi(4)}+ j'_{\pi(4)}=1$, and
$i'_{\pi(m)}\ne j'_{\pi(m)}$, i.e., $i'\in I_8(\tau)$.
\item [3)] The mapping $i\rightarrow i'$ is a one-to-one
mapping.  This together with the two facts above shows that $i'$
enumerates $I_8(\tau)$ as $i$ ranges over $I_8(\tau)$.

\item [4)]
$a_i-a_j-(a_{i'}-a_{j'})={H\over 2}(i_{\pi(1)}+j_{\pi(1)}
+i_{\pi(2)}+j_{\pi(2)}+i_{\pi(4)}+j_{\pi(4)}+i_{\pi(m)}+j_{\pi(m)})+2c_1(i_{\pi(3)}-j_{\pi(3)})={H\over
2}$. This indicates that $\xi^{a_i-a_j}+\xi^{a_{i'}-a_{j'}}=0$ for
any $i\in I_8(\tau)$.
\end{enumerate}
Hence equality \eqref{eq I8} holds.

Assume $i\in I_9(\tau)$, for convenience, we denote the six-tuple
$(i_{\pi(2)},j_{\pi(2)},i_{\pi(3)},j_{\pi(3)},i_{\pi(4)},j_{\pi(4)})\in
Z_2^6$ by $\mathcal {A}_1$, and
$(i'_{\pi(2)},j'_{\pi(2)},i'_{\pi(3)},j'_{\pi(3)},i'_{\pi(4)},j'_{\pi(4)})$
by $\mathcal {B}_1$.

Since $i_{\pi(2)}\ne j_{\pi(2)}$ and $i_{\pi(4)}\ne j_{\pi(4)}$, the
six-tuple $\mathcal {A}_1\in \Z_2^6$ has $16$ possibilities listed
in Table \ref{tab 11}. Note that $\pi(2)=2$, $\pi(3)=1$, $\pi(4)=3$,
the sign of $j-i$, will depend on the sign of value
$\Delta:=\sum_{k=2}^4(j_{\pi(k)}-i_{\pi(k)})2^{m-\pi(k)}$. If
$\Delta>0$, $\tau=j-i$; otherwise, $\tau=j+2^m-i$. When $\tau\in
\{k: 1\le k\le 2^{m-2}\}$, we have shown that in the $12$ cases, we
have $j>i+2^{m-2}\ge i+\tau$ or $j+2^m> i+2^{m-2}\ge i+\tau$. Hence,
the six-tuple $\mathcal {A}_1$ must be one of the following four
tuples: $(0,1,0,0,1,0)$, $(1,0,0,1,1,0)$, $(0,1,1,1,1,0)$, and
$(1,0,1,0,1,0)$. When $k\ne 2,3,4$, let $i'_{\pi(k)}= i_{\pi(k)}$
and $j'_{\pi(k)}=j_{\pi(k)}$. When $k=2,3,4$, $\mathcal {A}_1$ and
$\mathcal {B}_1$ are given in the Table \ref{tab 11}. We have the
following assertions.
\begin{enumerate}
\item [1)] $j'-i'\equiv j-i\equiv \tau~\bmod~2^m$ for any $i\in
I_9(\tau)$.
\item [2)] $i'$ satisfies $i'_{\pi(1)}\ne j'_{\pi(1)}$, $i'_{\pi(2)}\ne
j'_{\pi(2)}$, $i'_{\pi(4)}\ne j'_{\pi(4)}$, and $i'_{\pi(m)}\ne
j'_{\pi(m)}$, i.e., $i'\in I_9(\tau)$.
\item [3)] The mapping $i\rightarrow i'$ is a one-to-one
mapping.  This together with the two facts above indicates that $i'$
enumerates $I_9(\tau)$ as $i$ ranges over $I_9(\tau)$.

\item [4)]
$a_i-a_j-a_{i'}+a_{j'}={H\over 2}\sum_{k=1}^4(i_{\pi(k)}i_{\pi(k+1)}
-j_{\pi(k)}j_{\pi(k+1)}-i'_{\pi(k)}i'_{\pi(k+1)}+j'_{\pi(k)}j'_{\pi(k+1)})+\sum_{k=2}^4(i_{\pi(k)}
-j_{\pi(k)}-i'_{\pi(k)}+j'_{\pi(k)})=c_1-2c_2+{H\over 2}={H\over
2}$. This implies $\xi^{a_i-a_j}+\xi^{a_{i'}-a_{j'}}=0$ for any
$i\in I_9(\tau)$.
\end{enumerate}
Hence equality \eqref{eq I9} holds.

\begin{table}[ht]
\begin{center}
\caption{Values of $\mathcal {A}_1$ and their corresponding
$\mathcal {B}_1$}\label{tab 11}
\begin{tabular}{|c|c|c|c|c|}
  \hline
Item & $\mathcal {A}_1$ & $j-i$ & Remark & $\mathcal {B}_1$
\\ \hline \hline
$1$ & $(0,1,0,0,0,1)$ & $>0$ & $i+2^{m-2}<2^{m-3}+2^{m-2}\le j$  & \\
\hline
$2$ & $(0,1,0,1,0,1)$ & $>0$ & $i+2^{m-2}<2^{m-3}+2^{m-2}\le j$  & \\
\hline
$3$ & $(0,1,1,0,0,1)$ & $<0$ &  $i+2^{m-2}<(2^{m-1}+2^{m-3})+2^{m-2}<2^m+ j$  & \\
\hline
$4$ & $(0,1,1,1,0,1)$ & $>0$ & $i+2^{m-2}<2^{m-1}+2^{m-3}+2^{m-2}\le j$  & \\
\hline
$5$ & $(1,0,0,0,0,1)$ & $<0$ & $i+2^{m-2}<2^{m-1}+2^{m-2}<2^m+j$ & \\
\hline
$6$ & $(1,0,0,1,0,1)$ & $>0$ & $i+2^{m-2}<(2^{m-2}+2^{m-3})+2^{m-2}\le j$ & \\
\hline
$7$ & $(1,0,1,0,0,1)$ & $<0$ &  $\begin{array}{l}i+2^{m-2}<(2^{m-2}+2^{m-1}+2^{m-3})+2^{m-2}\\
=2^m+2^{m-3}\le 2^m+j\end{array}$ & \\
\hline
$8$ & $(1,0,1,1,0,1)$ & $<0$ & $\begin{array}{l}i+2^{m-2}<(2^{m-2}+2^{m-1}+2^{m-3})+2^{m-2}\\
<2^m+2^{m-1}+2^{m-3}\le 2^m+j\end{array}$  & \\
\hline

$9$ & $*(0,1,0,0,1,0)$ & $>0$ &    & $(1,0,0,1,1,0)$  \\
\hline
$10$ & $(0,1,0,1,1,0)$ & $>0$ & $i+2^{m-2}<2^{m-2}+2^{m-2}<2^{m-2}+2^{m-1}\le j$  & \\
\hline
$11$ & $(0,1,1,0,1,0)$ & $<0$ &  $i+2^{m-2}<(2^{m-1}+ 2^{m-2})+2^{m-2}\le 2^m+ j$ & \\
\hline
$12$ & $*(0,1,1,1,1,0)$ & $>0$ &   &  $(1,0,1,0,1,0)$ \\
\hline
$13$ & $(1,0,0,0,1,0)$ & $<0$ & $i+2^{m-2}<2^{m-2}+2^{m-2}\le 2^m+j$  & \\
\hline
$14$ & $*(1,0,0,1,1,0)$ & $>0$ &   &  $(0,1,0,0,1,0)$  \\
\hline
$15$ & $*(1,0,1,0,1,0)$ & $<0$ &   &  $(0,1,1,1,1,0)$  \\
\hline $16$ & $(1,0,1,1,1,0)$ & $<0$ & $i+2^{m-2}
<2^m+2^{m-1}\le 2^m+j$ & \\
\hline
\end{tabular}
\end{center}
\end{table}

By equalities \eqref{eq 2}-\eqref{eq I9}, the conclusion follows
immediately.

\done

\begin{lemma}\label{le 5}
Let $a$ be the sequence given by Definition \ref{de golay} and
satisfy the condition (A)-(3). Then we have $\sum_{i\in
I_3(\tau)}\xi^{a_i-a_j}=0$ for any $\tau\in \{k:1\le k\le
2^{m-2}\}$.
\end{lemma}

{\em Proof: }  We can partition $I_3(\tau)$ into the following two
disjoint subsets:
\begin{eqnarray*}
I_{10}(\tau)&=&\{0\le i\le 2^m-1: i_{\pi(1)}\ne j_{\pi(1)},
i_{\pi(3)}=
j_{\pi(3)}, i_{\pi(m)}\ne j_{\pi(m)}\};\\
I_{11}(\tau)&=&\{0\le i\le 2^m-1: i_{\pi(1)}\ne j_{\pi(1)},
i_{\pi(3)}\ne j_{\pi(3)}, i_{\pi(m)}\ne j_{\pi(m)}\}.
\end{eqnarray*}
Then $\sum_{i\in I_3(\tau)}\xi^{a_i-a_j}$ can be written as
\begin{eqnarray}\label{eq 5}\sum_{i\in I_3(\tau)}\xi^{a_i-a_j}=\sum_{i\in I_{10}(\tau)}\xi^{a_i-a_j}+\sum_{i\in
I_{11}(\tau)}\xi^{a_i-a_j}.\end{eqnarray}

Now we will show that
\begin{eqnarray}\label{eq I10}
\sum\limits_{i\in I_{10}(\tau)}\xi^{a_i-a_j}&=&0\\
\label{eq I11}\sum\limits_{i\in
I_{11}(\tau)}\xi^{a_i-a_j}&=&0\end{eqnarray} then $\sum_{i\in
I_3(\tau)}\xi^{a_i-a_j}=0$.

For any given $i\in I_{10}(\tau)$, let $i'$ and $j'$ be two integers
with binary representations defined by

\begin{eqnarray*}i'_{\pi(k)}=\left\{\begin{array}{ll} j_{\pi(k)}, & k=2\\
1-j_{\pi(k)}, & \mbox{otherwise} \end{array}\right.\end{eqnarray*}
and
\begin{eqnarray*}j'_{\pi(k)}=\left\{\begin{array}{ll} i_{\pi(k)}, & k=2\\
1-i_{\pi(k)}, & \mbox{otherwise.} \end{array}\right.\end{eqnarray*}
We have the following results.
\begin{enumerate}
\item [1)] $j'-i'\equiv j-i\equiv \tau~\bmod~2^m$ for any $i\in
I_{10}(\tau)$.
\item [2)] $i'$
satisfies $i'_{\pi(1)}\ne j'_{\pi(1)}$, $i'_{\pi(3)}= j'_{\pi(3)}$,
and $i'_{\pi(m)}\ne j'_{\pi(m)}$, i.e., $i'\in I_{10}(\tau)$.
\item [3)] The mapping $i\rightarrow i'$ is a one-to-one
mapping.  This together with the two facts above indicates that $i'$
enumerates $I_{10}(\tau)$ as $i$ ranges over $I_{10}(\tau)$.

\item [4)]
$a_i-a_j-(a_{i'}-a_{j'})={H\over
2}(i_{\pi(3)}+j_{\pi(3)}+i_{\pi(m)}+j_{\pi(m)})={H\over 2}$. This
implies $\xi^{a_i-a_j}+\xi^{a_{i'}-a_{j'}}=0$ for any $i\in
I_{10}(\tau)$.
\end{enumerate}
Hence equality \eqref{eq I10} holds.

Assume that $i\in I_{11}(\tau)$. For simplicity, we denote
$(i_{\pi(1)},j_{\pi(1)},i_{\pi(2)},j_{\pi(2)},i_{\pi(3)},j_{\pi(3)})$
by $\mathcal {A}_2$, and
$(i'_{\pi(1)},j'_{\pi(1)},i'_{\pi(2)},j'_{\pi(2)},i'_{\pi(3)},j'_{\pi(3)})$
by $\mathcal {B}_2$. Using the same argument as $i\in I_9(\tau)$,
the six-tuple $\mathcal {A}_2\in Z_2^6$ must be one of the four
following tuples: $(0,1,0,0,1,0)$, $(1,0,0,1,1,0)$, $(0,1,1,1,1,0)$,
and $(1,0,1,0,1,0)$.

When $k\ne 1,2,3$, let $i'_{\pi(k)}= i_{\pi(k)}$ and
$j'_{\pi(k)}=j_{\pi(k)}$. When $k=1,2,3$, $\mathcal {A}_2$ and
$\mathcal {B}_2$ are given in the Table \ref{tab 2}. We have the
following results.
\begin{enumerate}
\item [1)] $j'-i'\equiv j-i\equiv \tau~\bmod~2^m$ for any $i\in
I_{11}(\tau)$;
\item [2)] $i'$
satisfies $i'_{\pi(1)}\ne j'_{\pi(1)}$, $i'_{\pi(3)}\ne
j'_{\pi(3)}$, and $i'_{\pi(m)}\ne j'_{\pi(m)}$, i.e., $i'\in
I_{11}(\tau)$.
\item [3)] The mapping $i\rightarrow i'$ is a one-to-one mapping. This together with the two
 facts above indicates that $i'$ enumerates $I_{11}(\tau)$ as $i$ ranges over $I_{11}(\tau)$.

\item [4)]
$a_i-a_j-a_{i'}+a_{j'}={H\over 2}\sum_{k=1}^3(i_{\pi(k)}i_{\pi(k+1)}
-j_{\pi(k)}j_{\pi(k+1)}-i'_{\pi(k)}i'_{\pi(k+1)}+j'_{\pi(k)}j'_{\pi(k+1)})+\sum_{k=1}^3(i_{\pi(k)}
-j_{\pi(k)}-i'_{\pi(k)}+j'_{\pi(k)})=c_1-2c_2={H\over 2}$. This
implies $\xi^{a_i-a_j}+\xi^{a_{i'}-a_{j'}}=0$ for any $i\in
I_{11}(\tau)$.
\end{enumerate}
Hence equality \eqref{eq I11} holds.

\begin{table}[ht]
\begin{center}
\caption{Values of $\mathcal {A}_2$ and their corresponding
tuples $\mathcal {B}_2$}\label{tab 2}
\begin{tabular}{|c|c|c|}
  \hline
Item & $\mathcal {A}_2$ & $\mathcal {B}_2$
\\ \hline \hline
$1$ & $(0,1,0,0,1,0)$ & $(1,0,0,1,1,0)$  \\
 \hline
 $2$ & $(1,0,0,1,1,0)$ &  $(0,1,0,0,1,0)$  \\
 \hline
 $3$ & $(0,1,1,1,1,0)$ &  $(1,0,1,0,1,0)$ \\
 \hline
 $4$ & $(1,0,1,0,1,0)$  &  $(0,1,1,1,1,0)$  \\
  \hline
\end{tabular}
\end{center}
\end{table}

By \eqref{eq 5}, \eqref{eq I10} and \eqref{eq I11}, we finish the
proof.

\done

{\bf Proof of Theorem \ref{th 1}.} Note that
$R_a(\tau)=R_a(2^m-\tau)$ for any integer $\tau$. By equality
\eqref{eq auto}, it is sufficient to prove that
\begin{eqnarray*}\sum_{i\in I_3(\tau)}\xi^{a_i-a_j}=0,\,\,\,\,\tau\in \{k:1\le k\le 2^{m-2}\}\end{eqnarray*}
holds for the condition given by (A)-(1), (A)-(2) and (A)-(3). This
has been given in Lemmas \ref{le 3}, \ref{le 4} and \ref{le 5}.
Hence, the conclusion holds under the condition (A).

Define a mapping $\pi'(k)=\pi(m+1-k)$, $k\in \{1,\cdots,m\}$.
Replacing $\pi$ by $\pi'$, the conclusion under the condition (A')
follows immediately from the conclusion under the condition (A).

\done

\begin{remark}
The result in (A)-(1) of Theorem \ref{th 1} in the case of $H=2$ has
been reported in \cite{Gong}.
\end{remark}

{\bf Proof of Theorem \ref{th 2}.} Compared with the condition given
by (A)-(3) in Theorem \ref{th 1}, the condition given by (B) are the
same except for the value of $c_1-2c_2$. Because $c_1-2c_2$ is only
present in $i\in I_{11}(\tau)$, it is sufficient to prove
$\sum_{i\in I_{11}(\tau)}\xi^{a_i-a_j}=0$ to complete the proof for
this theorem.

Assume $i\in I_{11}(\tau)$. For simplicity to describe, we denote
$(i_{\pi(1)},j_{\pi(1)},i_{\pi(2)},j_{\pi(2)},i_{\pi(3)},j_{\pi(3)})$
by $\mathcal {A}_3$, and
$(i'_{\pi(1)},j'_{\pi(1)},i'_{\pi(2)},j'_{\pi(2)},i'_{\pi(3)},j'_{\pi(3)})$
by $\mathcal {B}_3$.

Since $i_{\pi(1)}\ne j_{\pi(1)}$, $i_{\pi(2)}\ne j_{\pi(2)}$, the
six-tuple $\mathcal {A}_3\in \Z_2^6$ has $16$ possibilities listed
in Table \ref{tab 3}. Note that $\pi(1)=2$, $\pi(2)=1$, $\pi(3)=3$,
the sign of $j-i$, will depend on the sign of the value
$\Delta:=\sum_{k=1}^3(j_{\pi(k)}-i_{\pi(k)})2^{m-\pi(k)}$. If
$\Delta>0$, $\tau=j-i$; otherwise, $\tau=j+2^m-i$. When $\tau\in
\{k: 2^{m-2}\le k\le 2^{m-1}\}$, we have shown that in the $12$
cases, we have $j>i+2^{m-1}\ge i+\tau$, $j+2^m>i+2^{m-1}\ge i+\tau$,
$i+\tau\ge i+2^{m+2}>j$, or $i+\tau\ge i+2^{m-2}>2^m+j$. Hence, the
six-tuple $\mathcal {A}_3$ must be one of the following four tuples:
$(0,1,0,0,1,0)$, $(1,0,0,1,1,0)$, $(0,1,1,1,1,0)$, and
$(1,0,1,0,1,0)$.

When $k\ne 1,2,3$, let $i'_{\pi(k)}= i_{\pi(k)}$ and
$j'_{\pi(k)}=j_{\pi(k)}$. When $k=1,2,3$, $\mathcal {A}_3$ and
$\mathcal {B}_3$ are given in the Table \ref{tab 3}. We have the
following assertions.
\begin{enumerate}
\item [1)] $j'-i'\equiv j-i\equiv \tau~\bmod~2^m$ for any $i\in
I_{11}(\tau)$.
\item [2)] $i'$
satisfies $i'_{\pi(1)}\ne j'_{\pi(1)}$, $i'_{\pi(3)}\ne
j'_{\pi(3)}$, and $i'_{\pi(m)}\ne  j'_{\pi(m)}$, i.e., $i'\in
I_{11}(\tau)$.
\item [3)] The mapping $i\rightarrow i'$ is a one-to-one mapping.
This together with the two facts above shows that $i'$ enumerates
$I_{11}(\tau)$ as $i$ ranges over $I_{11}(\tau)$.

\item [4)]
$a_i-a_j-a_{i'}+a_{j'}={H\over 2}\sum_{k=1}^3(i_{\pi(k)}i_{\pi(k+1)}
-j_{\pi(k)}j_{\pi(k+1)}-i'_{\pi(k)}i'_{\pi(k+1)}+j'_{\pi(k)}j'_{\pi(k+1)})+\sum_{k=1}^3(i_{\pi(k)}
-j_{\pi(k)}-i'_{\pi(k)}+j'_{\pi(k)})=c_1-2c_2+{H\over 2}={H\over
2}$. This implies $\xi^{a_i-a_j}+\xi^{a_{i'}-a_{j'}}=0$ for any
$i\in I_{11}(\tau)$.
\end{enumerate}
Hence  the equality \eqref{eq I11} holds.

\begin{table}[ht]
\begin{center}
\caption{Possibilities of $\mathcal {A}_3$ and their corresponding
tuples $\mathcal {B}_3$}\label{tab 3}\vspace{-4mm}
\begin{tabular}{|c|c|c|c|c|}
  \hline
Item & $\mathcal {A}_3$ & $j-i$ & Remark & $\mathcal {B}_3$
\\ \hline \hline
$1$ & $*(0,1,0,0,0,1)$ & $>0$ &  &  $(1, 0, 1, 0, 0, 1)$   \\
\hline
$2$ & $(0,1,0,1,0,1)$ & $>0$ & $i+2^{m-1}<2^{m-3}+2^{m-1}<j$  & \\
\hline
$3$ & $(0,1,1,0,0,1)$ & $<0$ &  $i+2^{m-1}<(2^{m-1}+2^{m-3})+2^{m-1}<2^m+ j$ & \\
\hline
$4$ & $*(0,1,1,1,0,1)$ & $>0$ &   &  $(1, 0, 0, 1, 0, 1)$ \\
\hline
$5$ & $(1,0,0,0,0,1)$ & $<0$ & $i+2^{m-1}<(2^{m-2}+2^{m-3})+2^{m-1}< j+2^m$ & \\
\hline
$6$ & $*(1,0,0,1,0,1)$ & $>0$ &  &  $(0, 1, 1, 1, 0, 1)$  \\
\hline
$7$ & $*(1,0,1,0,0,1)$ & $<0$ &  & $(0, 1, 0, 0, 0, 1)$  \\
\hline
$8$ & $(1,0,1,1,0,1)$ & $<0$ & $\begin{array}{l}i+2^{m-1}<(2^{m-2}+2^{m-1}+2^{m-3})+2^{m-1}\\
=2^m+2^{m-2}+2^{m-3}< 2^m+j\end{array}$ &  \\
\hline

$9$ & $(0,1,0,0,1,0)$ & $>0$ &  $i+2^{m-2}\ge 2^{m-3}+2^{m-2}>j$ &\\
\hline
$10$ & $(0,1,0,1,1,0)$ & $>0$ & $i+2^{m-1}<2\cdot 2^{m-3}+2^{m-1}=2^{m-2}+2^{m-1}\le j$  &\\
\hline
$11$ & $(0,1,1,0,1,0)$ & $<0$ &  $i+2^{m-1}<(2^{m-1}+2\cdot 2^{m-3})+2^{m-1}\le 2^m+ j$ & \\
\hline
$12$ & $(0,1,1,1,1,0)$ & $>0$ &  $i+2^{m-2}\ge (2^{m-1}+2^{m-3})+2^{m-2}>j$ & \\
\hline
$13$ & $(1,0,0,0,1,0)$ & $<0$ & $i+2^{m-1}<2^{m-1}+2^{m-1}\le 2^m+j$  & \\
\hline
$14$ & $(1,0,0,1,1,0)$ & $>0$ & $i+2^{m-1}<2^{m-1}+2^{m-1}\le 2^m+j$ &  \\
\hline
$15$ & $(1,0,1,0,1,0)$ & $<0$ &  $i+2^{m-2}\ge (2^{m-1}+2^{m-2}+2^{m-3})+2^{m-2}>2^m+j$ & \\
\hline
$16$ & $(1,0,1,1,1,0)$ & $<0$ & $i+2^{m-1}<2^m+2^{m-1}\le 2^m+j$ &  \\
\hline
\end{tabular}
\end{center}
\end{table}

Summarizing all cases above, the conclusion holds under the
condition (B).

Define a mapping $\pi'(k)=\pi(m+1-k)$, $k\in \{1,\cdots,m\}$.
Replacing $\pi$ by $\pi'$, the conclusion under the condition (B')
follows immediately from the conclusion under the condition (B).

\done

In order to prove Theorem \ref{th 3}, we need several lemmas.

\begin{lemma}\label{le 6}
Let $a$ be the sequence given by Definition \ref{de golay} and
satisfy the condition (C)-(1). Then one has $\sum_{i\in
I_3(\tau)}\xi^{a_i-a_j}=0$  for any $\tau\in \{k:1\le k\le
2^{m-3}\}\cup\{k:3\cdot 2^{m-3}\le k\le 2^{m-1}\}$.
\end{lemma}

{\em Proof:} We partition $I_3(\tau)$ into the following three
disjoint subsets:
\begin{eqnarray*}
I_{12}(\tau)&=&\{0\le i\le 2^m-1: i_{\pi(1)}\ne j_{\pi(1)},
i_{\pi(2)}= j_{\pi(2)}, i_{\pi(3)}\ne j_{\pi(3)},
i_{\pi(m)}\ne j_{\pi(m)}\};\\
I_{13}(\tau)&=&\{0\le i\le 2^m-1: i_{\pi(1)}\ne j_{\pi(1)},
i_{\pi(2)}=
j_{\pi(2)}, i_{\pi(3)}= j_{\pi(3)}, i_{\pi(m)}\ne j_{\pi(m)}\};\\
I_{14}(\tau)&=&\{0\le i\le 2^m-1: i_{\pi(1)}\ne j_{\pi(1)},
i_{\pi(2)}\ne j_{\pi(2)},i_{\pi(m)}\ne j_{\pi(m)}\}.
\end{eqnarray*}

First we will show that the subset $I_{12}(\tau)$ is an empty set.
In this case, $i_{\pi(1)}\ne j_{\pi(1)}$ and $i_{\pi(3)}\ne
j_{\pi(3)}$ implies $(i_{\pi(1)},j_{\pi(1)},i_{\pi(3)},j_{\pi(3)})$
must be one of $(0,1,0,1)$, $(0,1,1,0)$, $(1,0,0,1)$, and
$(1,0,1,0)$.  When $\tau\in \{k:1\le k\le 2^{m-3}\}\cup\{k:3\cdot
2^{m-3}\le k\le 2^{m-1}\}$, we have shown that in Tables \ref{tab 4}
and \ref{tab 5}, we have $j>i+\tau$, $j+2^m>i+\tau$, $i+\tau>j$, or
$i+\tau>2^m+j$. This contradicts with $j\equiv (i+\tau)~\bmod~2^m$.

\begin{table}[ht]
\begin{center}
\caption{The case $\tau\in \{k:1\le k\le 2^{m-3}\}$}\label{tab 4}
\begin{tabular}{|c|c|c|c|}
  \hline
Item & $(i_{\pi(1)},j_{\pi(1)},i_{\pi(3)},j_{\pi(3)})$ & $j-i$ &
Remark
\\ \hline \hline
$1$ & $(0,1,0,1)$ & $>0$ &
$i+\tau<2^{m-2}+2^{m-1}=j_{\pi(3)}2^{m-2}+j_{\pi(1)}2^{ m-1}\le
j$   \\
\hline $2$ & $(0,1,1,0)$ & $>0$ & $\begin{array}{l}
i+\tau<(i_{\pi(2)}2^{m-3}+2^{m-2}+2^{m-3})+2^{m-3}\\
=i_{\pi(2)}2^{ m-3}+2^{m-1}=j_{\pi(2)}2^{ m-3}+2^{m-1}\le
j\end{array}$  \\
\hline $3$ & $(1,0,0,1)$ & $<0$ &  $\begin{array}{l}
i+\tau<(i_{\pi(1)}2^{m-1}+i_{\pi(2)}2^{m-3}+2^{m-3})+2^{m-1}\\
<j_{\pi(2)}2^{ m-3}+j_{\pi(3)}2^{
m-2}+2^m\le j+2^m\end{array}$  \\
\hline $4$ & $(1,0,1,0)$ & $>0$ &  $\begin{array}{l}
i+\tau<(i_{\pi(1)}2^{m-1}+i_{\pi(2)}2^{m-3}+2^{m-2}+2^{m-3})\\
+2^{m-3}=j_{\pi(2)}2^{ m-2}+2^m\le
j+2^m\end{array}$ \\
\hline
\end{tabular}
\end{center}
\end{table}

\begin{table}[ht]
\begin{center}
\caption{The case $\tau\in \{k:3\cdot 2^{m-3}\le k\le
2^{m-1}\}$\label{tab 5}}
\begin{tabular}{|c|c|c|c|}
  \hline
Item & $(i_{\pi(1)},j_{\pi(1)},i_{\pi(3)},j_{\pi(3)})$ & $j-i$ &
Remark
\\ \hline \hline
$1$ & $(0,1,0,1)$ & $>0$ &
$i+\tau<2^{m-2}+2^{m-1}=j_{\pi(3)}2^{m-2}+j_{\pi(1)}2^{ m-1}\le
j$   \\
\hline $2$ & $(0,1,1,0)$ & $>0$ & $\begin{array}{l}
i+\tau\ge(i_{\pi(2)}2^{m-3}+2^{m-2})+3\cdot 2^{m-3}\\
=j_{\pi(2)}2^{ m-3}+2^{m-1}+2^{m-3}>j\end{array}$  \\
\hline $3$ & $(1,0,0,1)$ & $<0$ &  $\begin{array}{l}
i+\tau<(i_{\pi(1)}2^{m-1}+i_{\pi(2)}2^{m-3}+2^{m-3})+2^{m-1}\\
<j_{\pi(2)}2^{ m-3}+j_{\pi(3)}2^{
m-2}+2^m\le j+2^m\end{array}$  \\
\hline $4$ & $(1,0,1,0)$ & $>0$ &  $\begin{array}{l}
i+\tau\ge(i_{\pi(1)}2^{m-1}+i_{\pi(2)}2^{m-3}+2^{m-2})+3\cdot 2^{m-3}\\
=j_{\pi(2)}2^{
m-2}+2^m+2^{m-3}> j+2^m\end{array}$ \\
\hline
\end{tabular}
\end{center}
\end{table}

By the discussion above, the set $I_{12}(\tau)$ is an empty set.
Then $\sum_{i\in I_3(\tau)}\xi^{a_i-a_j}$ can be written as
\begin{eqnarray}\label{eq 4}\sum_{i\in
I_3(\tau)}\xi^{a_i-a_j}=\sum_{i\in
I_{13}(\tau)}\xi^{a_i-a_j}+\sum_{i\in
I_{14}(\tau)}\xi^{a_i-a_j}.\end{eqnarray}

Now we will show that
\begin{eqnarray}\label{eq I13}
\sum\limits_{i\in I_{13}(\tau)}\xi^{a_i-a_j}&=&0\\
\label{eq I14}\sum\limits_{i\in
I_{14}(\tau)}\xi^{a_i-a_j}&=&0\end{eqnarray} then $\sum_{i\in
I_3(\tau)}\xi^{a_i-a_j}=0$.

For the case $i\in I_{13}(\tau)$, let $i'$ and $j'$ be two integers
with binary representations defined by
\begin{eqnarray*}i'_{\pi(k)}=\left\{\begin{array}{ll} 1-i_{\pi(k)}, & k=2\\
i_{\pi(k)}, & k\ne 2\end{array}\right.\end{eqnarray*} and
\begin{eqnarray*}j'_{\pi(k)}=\left\{\begin{array}{ll} 1-j_{\pi(k)}, & k=2\\
j_{\pi(k)}, & k\ne 2.\end{array}\right.\end{eqnarray*} We can derive
the following results.
\begin{enumerate}
\item [1)] $j'-i'\equiv j-i\equiv \tau~\bmod~2^m$ for any $i\in
I_{13}(\tau)$ by using $i_{\pi(2)}= j_{\pi(2)}$.
\item [2)] $i'$
satisfies $i'_{\pi(1)}\ne j'_{\pi(1)}$, $i'_{\pi(2)}= j'_{\pi(2)}$,
$i'_{\pi(3)}= j'_{\pi(3)}$, and $i'_{\pi(m)}\ne j'_{\pi(m)}$, i.e.,
$i'\in I_{13}(\tau)$.
\item [3)] The mapping $i\rightarrow i'$ is a one-to-one mapping.
This together with the two facts above shows that $i'$ enumerates
$I_{13}(\tau)$ as $i$ ranges over $I_{13}(\tau)$.

\item [4)]
$a_i-a_j-(a_{i'}-a_{j'})={H\over
2}(i_{\pi(1)}+j_{\pi(1)}+i_{\pi(3)}+j_{\pi(3)})={H\over 2}$. This
implies $\xi^{a_i-a_j}+\xi^{a_{i'}-a_{j'}}=0$ for any $i\in
I_{13}(\tau)$.
\end{enumerate}
Hence equality \eqref{eq I13} holds.

For the case $i\in I_{14}(\tau)$, let $i'$ and $j'$ be two integers
with binary representations defined by
\begin{eqnarray*}i'_{\pi(k)}=\left\{\begin{array}{ll} 1-i_{\pi(k)}, & k=1\\
i_{\pi(k)}, & k\ne 1\end{array}\right.\end{eqnarray*} and
\begin{eqnarray*}j'_{\pi(k)}=\left\{\begin{array}{ll} 1-j_{\pi(k)}, & k=1\\
j_{\pi(k)}, & k\ne 1.\end{array}\right.\end{eqnarray*}

We can derive the following assertions.
\begin{enumerate}
\item [1)] $j'-i'\equiv j-i\equiv \tau~\bmod~2^m$ for any $i\in
I_{14}(\tau)$.
\item [2)] $i'$
satisfies $i'_{\pi(1)}\ne j'_{\pi(1)}$, $i'_{\pi(2)}\ne
j'_{\pi(2)}$, and $i'_{\pi(m)}\ne j'_{\pi(m)}$, i.e., $i'\in
I_{14}(\tau)$.
\item [3)] The mapping $i\rightarrow i'$ is a one-to-one mapping.
This together with the two facts above indicates that $i'$
enumerates $I_{14}(\tau)$ as $i$ ranges over $I_{14}(\tau)$.

\item [4)]
$a_i-a_j-(a_{i'}-a_{j'})={H\over 2}(i_{\pi(2)}+j_{\pi(2)})={H\over
2}$. This implies $\xi^{a_i-a_j}+\xi^{a_{i'}-a_{j'}}=0$ for any
$i\in I_{14}(\tau)$.
\end{enumerate}
Hence equality \eqref{eq I14} holds.


\done

\begin{lemma}\label{le 7}
Let $a$ be the sequence given by Definition \ref{de golay} and
satisfy the condition (C)-(2). Then one has $\sum_{i\in
I_3(\tau)}\xi^{a_i-a_j}=0$  for any $\tau\in \{k:1\le k\le
2^{m-3}\}\cup\{k:3\cdot 2^{m-3}\le k\le 2^{m-1}\}$.
\end{lemma}

{\em Proof: } The set $I_3(\tau)$ is divided into two disjoint
subsets $I_{14}(\tau)$ and $I_{15}(\tau)$, where
\begin{eqnarray*}
I_{15}(\tau)&=&\{0\le i\le 2^m-1: i_{\pi(1)}\ne j_{\pi(1)},
i_{\pi(2)}= j_{\pi(2)}, i_{\pi(m)}\ne j_{\pi(m)}\}.
\end{eqnarray*}

Using the same argument as $i\in I_{12}(\tau)$ and $i\in
I_{14}(\tau)$, we have that the subset $I_{15}(\tau)$ is an empty
set and $\sum_{i\in I_{14}(\tau)}\xi^{a_i-a_j}=0$. We also have
$\sum_{i\in I_3(\tau)}\xi^{a_i-a_j}=\sum_{i\in
I_{14}(\tau)}\xi^{a_i-a_j}=0$.

\done

\begin{lemma}\label{le 8}
Let $a$ be the sequence given by Definition \ref{de golay} and
satisfy the condition (C)-(3). Then one has $\sum_{i\in
I_3(\tau)}\xi^{a_i-a_j}=0$ where $$\tau\in \{k:1\le k\le
2^{m-3}\}\cup\{k:3\cdot 2^{m-3}\le k\le 2^{m-1}\}.$$
\end{lemma}

{\em Proof: } For simplicity, we denote
$(i_{\pi(1)},j_{\pi(1)},i_{\pi(2)},j_{\pi(2)},i_{\pi(3)},j_{\pi(3)},i_{\pi(4)},j_{\pi(4)})$
by $\mathcal {A}_4$ and the corresponding eight-tuple
$(i'_{\pi(1)},j'_{\pi(1)},i'_{\pi(2)}$,
$j'_{\pi(2)},i'_{\pi(3)},j'_{\pi(3)},i'_{\pi(4)},j'_{\pi(4)})$ by
$\mathcal {B}_4$.

Assume that $i\in I_3(\tau)$. The eight-tuple $\mathcal {A}_4\in
\Z_2^8$ has $128$ possibilities since $i_{\pi(1)}\ne j_{\pi(1)}$.
Note that $\pi(1)=2$, $\pi(2)=4$, $\pi(3)=1$, $\pi(4)=3$, the sign
of $j-i$ will depend on the sign of the value
$\Delta:=\sum_{k=1}^4(j_{\pi(k)}-i_{\pi(k)})2^{m-\pi(k)}$. If
$\Delta>0$, $\tau=j-i$; otherwise, $\tau=j+2^m-i$. When $\tau\in
\{k:1\le k\le 2^{m-3}\}\cup\{k:3\cdot 2^{m-3}\le k\le 2^{m-1}\}$,
there are $104$ possible pairs $(i,j)$ that satisfy $j>i+2^{m-1}\ge
i+\tau$, $j+2^m>i+2^{m-1}\ge i+\tau$, $i+\tau\ge i+2^{m+2}>j$, or
$i+\tau\ge i+2^{m-2}>2^m+j$. Hence, the eight-tuple $\mathcal
{A}_4\in Z_2^8$ must be one of the remaining 24 possibilities as
shown in Table \ref{tab 7}. When $k=1,2,3,4$, the corresponding
tuples $\mathcal {B}_4$ are also given for any given $\mathcal
{A}_4$. When $k>4$, let $i'_{\pi(k)}= i_{\pi(k)}$ and
$j'_{\pi(k)}=j_{\pi(k)}$.

\begin{table}[ht]
\begin{center}
\caption{Possibilities of $\mathcal {A}_4$ and their corresponding
tuples $\mathcal {B}_4$}\label{tab 7}
\begin{tabular}{|c|c|c||c|c|c|}
  \hline
Item  & $\mathcal {A}_4$  & $\mathcal {B}_4$ & Item  & $\mathcal
{A}_4$  & $\mathcal {B}_4$
\\ \hline \hline
$1$ & $(0,1,1,0,1,1,1,0)$ & $(1,0,1,0,1,0,1,0)$ &　$13$ & $(1,0,1,0,1,0,1,0)$ & $(0,1,1,0,1,1,1,0)$ \\
\hline
$2$ & $(0,1,1,0,0,0,1,0)$ & $(1,0,1,0,0,1,1,0)$ &　$14$ & $(1,0,1,0,0,1,1,0)$ & $(0,1,1,0,0,0,1,0)$ \\
\hline
$3$ & $(0,1,0,0,1,1,1,0)$ & $(1,0,0,0,0,1,1,0)$ &　$15$ & $(1,0,0,0,0,1,1,0)$ & $(0,1,0,0,1,1,1,0)$ \\
\hline
$4$ & $(0,1,0,0,0,0,1,0)$ & $(1,0,0,0,1,0,1,0)$ &　$16$ & $(1,0,0,0,1,0,1,0)$ & $(0,1,0,0,0,0,1,0)$ \\
\hline
$5$ & $(0,1,1,1,0,0,1,0)$ & $(1,0,1,1,0,1,1,0)$ &　$17$ & $(1,0,1,1,0,1,1,0)$ & $(0,1,1,1,0,0,1,0)$ \\
\hline
$6$ & $(0,1,0,0,0,0,0,1)$ & $(1,0,0,0,0,1,0,1)$ &　$18$ & $(1,0,0,0,0,1,0,1)$ & $(0,1,0,0,0,0,0,1)$ \\
\hline
$7$ & $(0,1,0,0,1,1,0,1)$ & $(1,0,0,0,1,0,0,1)$ &　$19$ & $(1,0,0,0,1,0,0,1)$ & $(0,1,0,0,1,1,0,1)$ \\
\hline
$8$ & $(0,1,0,1,0,0,0,1)$ & $(1,0,0,1,1,0,0,1)$ &　$20$ & $(1,0,0,1,1,0,0,1)$ & $(0,1,0,1,0,0,0,1)$ \\
\hline
$9$ & $(0,1,0,1,1,1,0,1)$ & $(1,0,0,1,0,1,0,1)$ &　$21$ & $(1,0,0,1,0,1,0,1)$ & $(0,1,0,1,1,1,0,1)$ \\
\hline
$10$ & $(0,1,1,1,0,0,0,1)$ & $(1,0,1,1,1,0,0,1)$ &　$22$ & $(1,0,1,1,1,0,0,1)$ & $(0,1,1,1,0,0,0,1)$ \\
\hline
$11$ & $(0,1,1,1,1,1,0,1)$ &  $(1,0,1,1,0,1,0,1)$ &　$23$ & $(1,0,1,1,0,1,0,1)$  & $(0,1,1,1,1,1,0,1)$ \\
\hline
$12$ & $(0,1,1,1,1,1,1,0)$ & $(1,0,1,1,1,0,1,0)$ &　$24$ & $(1,0,1,1,1,0,1,0)$ & $(0,1,1,1,1,1,1,0)$ \\
\hline
\end{tabular}
\end{center}
\end{table}

We have the following assertions.
\begin{enumerate}
\item [1)] $j'-i'\equiv j-i\equiv \tau~\bmod~2^m$ for any $i\in
I_3(\tau)$.
\item [2)] $i'$
satisfies $i'_{\pi(1)}\ne j'_{\pi(1)}$, and $i'_{\pi(m)}\ne
j'_{\pi(m)}$, i.e., $i'\in I_3(\tau)$.
\item [3)] The mapping $i\rightarrow i'$ is a one-to-one
mapping.  This together with the two facts above indicates that $i'$
enumerates $I_3(\tau)$ as $i$ ranges over $I_3(\tau)$.

\item [4)]
$a_i-a_j-a_{i'}+a_{j'}={H\over 2}\sum_{k=1}^3(i_{\pi(k)}i_{\pi(k+1)}
-j_{\pi(k)}j_{\pi(k+1)}-i'_{\pi(k)}i'_{\pi(k+1)}+j'_{\pi(k)}j'_{\pi(k+1)})+\sum_{k=1}^3(i_{\pi(k)}
-j_{\pi(k)}-i'_{\pi(k)}+j'_{\pi(k)})=c_1-2c_2+{H\over 2}={H\over
2}$. This implies $\xi^{a_i-a_j}+\xi^{a_{i'}-a_{j'}}=0$ for any
$i\in I_3(\tau)$.
\end{enumerate}
Hence we have $\sum_{i\in I_3(\tau)}\xi^{a_i-a_j}=0$.

\done

\begin{lemma}\label{le 9}
Let $a$ be the sequence given by Definition \ref{de golay} and
satisfy the condition (C)-(4). Then one has $\sum_{i\in
I_3(\tau)}\xi^{a_i-a_j}=0$ for any $\tau\in \{k:1\le k\le
2^{m-3}\}\cup\{k:3\cdot 2^{m-3}\le k\le 2^{m-1}\}$.
\end{lemma}

{\em Proof: } For simplicity, we denote
$(i_{\pi(1)},j_{\pi(1)},i_{\pi(2)},j_{\pi(2)},i_{\pi(3)},j_{\pi(3)},i_{\pi(4)},j_{\pi(4)})$
by $\mathcal {A}_5$ and the corresponding eight-tuple
$(i'_{\pi(1)},j'_{\pi(1)},i'_{\pi(2)}$,
$j'_{\pi(2)},i'_{\pi(3)},j'_{\pi(3)},i'_{\pi(4)},j'_{\pi(4)})$ by
$\mathcal {B}_5$.

Assume $i\in I_3(\tau)$, the eight-tuple $\mathcal {A}_5\in \Z_2^8$
has $128$ possibilities since $i_{\pi(1)}\ne j_{\pi(1)}$. Note that
$\pi(1)=2$, $\pi(2)=3$, $\pi(3)=1$, $\pi(4)=4$, the sign of $j-i$
will depend on the sign of the value
$\Delta:=\sum_{k=1}^4(j_{\pi(k)}-i_{\pi(k)})2^{m-\pi(k)}$. If
$\Delta>0$, $\tau=j-i$; otherwise, $\tau=j+2^m-i$. When $\tau\in
\{k:1\le k\le 2^{m-3}\}\cup\{k:3\cdot 2^{m-3}\le k\le 2^{m-1}\}$,
there are $104$ pairs $(i,j)$ that satisfy $j>i+2^{m-1}\ge i+\tau$,
$j+2^m>i+2^{m-1}\ge i+\tau$, $i+\tau\ge i+2^{m+2}>j$, or $i+\tau\ge
i+2^{m-2}>2^m+j$. Hence, the eight-tuple $\mathcal {A}_5\in Z_2^8$
must be one of the left 24 pairs in Table \ref{tab 8}. When
$k=1,2,3,4$, the corresponding tuples $\mathcal {B}_5$ are also
given for any given $\mathcal {A}_5$. When $k>4$, let $i'_{\pi(k)}=
i_{\pi(k)}$ and $j'_{\pi(k)}=j_{\pi(k)}$.

\begin{table}[ht]
\begin{center}
\caption{Possibilities of $\mathcal {A}_5$ and their corresponding
tuples $\mathcal {B}_5$}\label{tab 8}
\begin{tabular}{|c|c|c||c|c|c|}
  \hline
Item  & $\mathcal {A}_5$  & $\mathcal {B}_5$ & Item  & $\mathcal
{A}_5$  & $\mathcal {B}_5$
\\ \hline\hline
$1$ & $(0,1,1,0,1,1,1,0)$ & $(1,0,1,0,1,0,1,0)$ &　$13$ & $(1,0,1,0,1,0,1,0)$ & $(0,1,1,0,1,1,1,0)$ \\
\hline
$2$ & $(0,1,1,0,0,0,1,0)$ & $(1,0,1,0,0,1,1,0)$ &　$14$ & $(1,0,1,0,0,1,1,0)$ & $(0,1,1,0,0,0,1,0)$ \\
\hline
$3$ & $(0,1,0,1,0,0,0,0)$ & $(1,0,0,1,1,0,0,0)$ &　$15$ & $(1,0,0,1,1,0,0,0)$ & $(0,1,0,1,0,0,0,0)$ \\
\hline
$4$ & $(0,1,0,1,0,0,1,1)$ & $(1,0,0,1,0,1,1,1)$ &　$16$ & $(1,0,0,1,0,1,1,1)$ & $(0,1,0,1,0,0,1,1)$ \\
\hline
$5$ & $(0,1,0,1,1,1,0,0)$ & $(1,0,0,1,0,1,0,0)$ &　$17$ & $(1,0,0,1,0,1,0,0)$ & $(0,1,0,1,1,1,0,0)$ \\
\hline
$6$ & $(0,1,0,1,1,1,1,1)$ & $(1,0,0,1,1,0,1,1)$ &　$18$ & $(1,0,0,1,1,0,1,1)$ & $(0,1,0,1,1,1,1,1)$ \\
\hline
$7$ & $(0,1,1,0,0,0,0,0)$ & $(1,0,1,0,0,1,0,0)$ &　$19$ & $(1,0,1,0,0,1,0,0)$ & $(0,1,1,0,0,0,0,0)$ \\
\hline
$8$ & $(0,1,0,1,0,0,0,1)$ & $(1,0,0,1,1,0,0,1)$ &　$20$ & $(1,0,0,1,1,0,0,1)$ & $(0,1,0,1,0,0,0,1)$ \\
\hline
$9$ & $(0,1,0,1,1,1,0,1)$ & $(1,0,0,1,0,1,0,1)$ &　$21$ & $(1,0,0,1,0,1,0,1)$ & $(0,1,0,1,1,1,0,1)$ \\
\hline
$10$ & $(0,1,1,0,0,0,1,1)$ & $(0,1,1,0,1,1,1,1)$  &　$22$ &  $(0,1,1,0,1,1,1,1)$ & $(0,1,1,0,0,0,1,1)$ \\
\hline
$11$ & $(0,1,1,0,1,1,0,0)$ & $(1,0,1,0,1,0,0,0)$  &　$23$ & $(1,0,1,0,1,0,0,0)$  & $(0,1,1,0,1,1,0,0)$ \\
\hline
$12$ & $(1,0,1,0,1,0,1,1)$ & $(1,0,1,0,0,1,1,1)$ &　$24$ & $(1,0,1,0,0,1,1,1)$ & $(1,0,1,0,1,0,1,1)$ \\
\hline
\end{tabular}
\end{center}
\end{table}

We have the following assertions.
\begin{enumerate}
\item [1)] $j'-i'\equiv j-i\equiv \tau~\bmod~2^m$ for any $i\in
I_3(\tau)$.
\item [2)] $i'$
satisfies $i'_{\pi(1)}\ne j'_{\pi(1)}$, and $i'_{\pi(m)}\ne
j'_{\pi(m)}$, i.e., $i'\in I_3(\tau)$.
\item [3)] The mapping $i\rightarrow i'$ is a one-to-one
mapping.  This together with the two facts above indicates that $i'$
enumerates $I_3(\tau)$ as $i$ ranges over $I_3(\tau)$.

\item [4)]
$a_i-a_j-a_{i'}+a_{j'}={H\over 2}\sum_{k=1}^3(i_{\pi(k)}i_{\pi(k+1)}
-j_{\pi(k)}j_{\pi(k+1)}-i'_{\pi(k)}i'_{\pi(k+1)}+j'_{\pi(k)}j'_{\pi(k+1)})+\sum_{k=1}^3(i_{\pi(k)}
-j_{\pi(k)}-i'_{\pi(k)}+j'_{\pi(k)})=c_1-2c_2+{H\over 2}={H\over
2}$. This implies $\xi^{a_i-a_j}+\xi^{a_{i'}-a_{j'}}=0$ for any
$i\in I_3(\tau)$.
\end{enumerate}
Hence we have $\sum_{i\in I_3(\tau)}\xi^{a_i-a_j}=0$.

\done

{\bf Proof of Theorem \ref{th 3}. } Since $R_a(\tau)=R_a(2^m-\tau)$
for any $\tau$, then by \eqref{eq auto}, it is sufficient to prove
\begin{eqnarray*}\sum_{i\in I_3(\tau)}\xi^{a_i-a_j}=0\end{eqnarray*}
is equal to zero for any $\tau\in \{k:1\le k\le
2^{m-3}\}\cup\{k:3\cdot 2^{m-3}\le k\le 2^{m-1}\}$, which have been
given in Lemmas \ref{le 6}-\ref{le 9} for the condition (C)-(1),
(C)-(2), (C)-(3) and (C)-(4). Hence, the conclusion holds under the
condition (C).

Define a mapping $\pi'(k)=\pi(m+1-k)$, $k\in \{1,\cdots,m\}$.
Replacing $\pi$ by $\pi'$, the conclusion under the condition (C')
follows immediately from the conclusion under the condition (C).

\done

\section{Zero Autocorrelation Zone of $4^q$-QAM Golay Complementary Sequences}

In this section, we will consider the ZACZ of $4^q$-QAM Golay
complementary sequences defined by \eqref{eq qam}, which are based
on the quaternary Golay sequences. So throughout this section, we
always assume that $H=4$ and $\xi$ is the primitive $4$-th root of
unity. For convenience to describe, denote $s_{i,0}:=0$ for $0\le i
<2^m$.

\subsection{Results}

\begin{theorem}\label{th 4}
If the $4^q$-QAM Golay complementary sequence $A$, defined by
\eqref{eq qam} with ($s_{i,e}=d_{e,0}+d_{e,1}i_{\pi(m)}$ or
$s_{i,e}=d_{e,0}+d_{e,1}i_{\pi(1)}$) for any $d_{e,0},d_{e,1}\in
\Z_4$, satisfies one of the condition listed in (A) or (A'), then
the sequence $A$ has the following property:
$$R_A(\tau)=0,\,\,\,\,\,\,\, \tau\in (0,2^{m-2}]\cup [3\cdot
2^{m-2},2^m).$$ In other words, in one period $[0,2^m)$, it has two
zero autocorrelation zones of length $2^{m-2}$, given by
$(0,2^{m-2}]$ and $[3\cdot 2^{m-2},2^m)$.
\end{theorem}

\begin{theorem}\label{th 5}
If the $4^q$-QAM Golay complementary sequence $A$, defined by
\eqref{eq qam} with ($s_{i,e}=d_{e,0}+d_{e,1}i_{\pi(m)}$ or
$s_{i,e}=d_{e,0}+d_{e,1}i_{\pi(1)}$) for any $d_{e,0},d_{e,1}\in
\Z_4$, satisfies one of the condition listed in (B) or (B'), then
the sequence $A$ has the following property:
$$R_A(\tau)=0, \,\,\,\,\,\,\, \tau\in [2^{m-2},3\cdot 2^{m-2}].$$ In
other words, in one period $[0,2^m)$, it has a zero autocorrelation
zone of length $2^{m-1}+1$, given by $[2^{m-2},3\cdot 2^{m-2}]$.
\end{theorem}

\begin{theorem}\label{th 6}
If the $4^q$-QAM Golay complementary sequence $A$, defined by
\eqref{eq qam} with ($s_{i,e}=d_{e,0}+d_{e,1}i_{\pi(m)}$ or
$s_{i,e}=d_{e,0}+d_{e,1}i_{\pi(1)}$) for any $d_{e,0},d_{e,1}\in
\Z_4$, satisfies one of the condition listed in (C) or (C'), then
the sequence $A$ has the following property:
$$R_A(\tau)=0, \,\,\,\,\,\tau\in (0,2^{m-3}]\cup [3\cdot 2^{m-3},
5\cdot 3^{m-3}]\cup [7\cdot 2^{m-3},2^m).$$ In other words, in one
period $[0,2^m)$, it has three zero autocorrelation zones of
respective length $2^{m-3}$, $2^{m-2}+1$, $2^{m-3}$, given by
$(0,2^{m-3}]$, $[3\cdot 2^{m-3}, 5\cdot 3^{m-3}]$ and $[7\cdot
2^{m-3},2^m)$.
\end{theorem}

\subsection{Proofs of the Results}

In Section $3$, the idea of the proof on the ZACZ of Golay sequence
$a$ is to define one-to-one mappings $i\rightarrow i'$ and
$j\rightarrow j'$, $0\le i\le 2^m-1$ such that
$\xi^{a_i-a_j}+\xi^{a_{i'}-a_{j'}}=0$. Hence we have
$$2\sum_{i=0}^{2^m-1}\xi^{a_i-a_j}=\sum_{i=0}^{2^m-1}\xi^{a_i-a_j}+\sum_{i'=0}^{2^m-1}\xi^{a_{i'}-a_{j'}}
=\sum_{i=0}^{2^m-1}\left(\xi^{a_i-a_j}+\xi^{a_{i'}-a_{j'}}\right)=0.$$
That is $\sum_{i=0}^{2^m-1}\xi^{a_i-a_j}=0$.

Note that $a_{i,0}$ is a quaternary Golay sequence. Under those
definitions of $(i',j')$ and conditions of $\pi$ and $(c_1,c_2)$ in
Section $3$, we have
\begin{eqnarray}\label{eq a0}a_{i,0}-a_{j,0}-(a_{i',0}-a_{j',0})=2\end{eqnarray} and
$$i_{\pi(m)}=i'_{\pi(m)},j_{\pi(m)}=j'_{\pi(m)}.$$
Let $s_{i,k}=d_{0,k}+d_{1,k}i_{\pi(m)}$, $1\le k\le q-1$, then the
latter equality indicates that
\begin{eqnarray}\label{eq
case1}s_{i,e}-s_{j,f}=s_{i',e}-s_{j',f}\end{eqnarray} for any $0\le
e,f\le q-1$. Equalities \eqref{eq a0} and \eqref{eq case1} implies
that
\begin{eqnarray*}
a_{i,e}-a_{j,f}-(a_{i',e}-a_{j',f})=
a_{i,0}-a_{j,0}-(a_{i',0}-a_{j',0})+s_{i,e}-s_{j,f}-(s_{i',e}-s_{j',f})=2
\end{eqnarray*}
or
$$\xi^{a_{i,e}-a_{j,f}}=\xi^{a_{i',e}-a_{j',f}}.$$
Similar to the discussion to the Golay sequence in Section 3, we
have
$$2\sum_{i=0}^{2^m-1}\xi^{a_{i,e}-a_{j,f}}=
\sum_{i=0}^{2^m-1}\xi^{a_{i,e}-a_{j,f}}+\sum_{i'=0}^{2^m-1}\xi^{a_{i',e}-a_{j',f}}
=\sum_{i=0}^{2^m-1}\left(\xi^{a_{i,e}-a_{j,f}}+\xi^{a_{i',e}-a_{j',f}}\right)=0$$
for any $0\le e,f\le q-1$, i.e.,
$\sum_{i=0}^{2^m-1}\xi^{a_{i,e}-a_{j,f}}=0$. Hence, for any given
$\tau$, $1\le \tau\le 2^m-1$,
\begin{eqnarray*}
R_A(\tau)
&=&\sum_{i=0}^{2^m-1}\left(\gamma\sum_{e=0}^{q-1}r_e\xi^{a_{i,e}}\right)\left(\gamma\sum_{f=0}^{q-1}r_f\xi^{a_{j,f}}\right)^*\\
&=&\sum_{i=0}^{2^m-1}\sum_{e,f=0}^{q-1}r_er_f\xi^{a_{i,e}-a_{j,f}}\\
&=&\sum_{e,f=0}^{q-1}r_er_f\sum_{i=0}^{2^m-1}\xi^{a_{i,e}-a_{j,f}}\\
&=&0.
\end{eqnarray*}

Naturally, the proofs of Theorems \ref{th 4}, \ref{th 5} and \ref{th
6} are similar to the proofs of Theorems \ref{th 1}, \ref{th 2} and
\ref{th 3}, respectively. So we omit them here.

We have presented the ZACZ for certain QAM Golay complementary
sequences in Cases 1 and 2 in Fact \ref{th Li}. For the QAM Golay
sequences in Case 3 under the conditions in Theorems \ref{th 4} -
\ref{th 6} as above, some have a large ZACZ, while others do not.
The following three examples under the condition in (A)-(1), (A)-(2)
and (A)-(3) of Theorem \ref{th 4} illustrate this fact. The first
sequence has a ZACZ of length $8$, while the other two do not have.

\begin{example}
Let $q=2$ and $m=5$. Let $\pi=(1)$, $c_1=0$ and
$s_i^{(1)}=1+i_{\pi(2)}+i_{\pi(3)}$. Then such $16$-QAM Golay
sequence $A$ defined in Theorem \ref{th Li} has $R_A(\tau)=0$ for
$\tau\in (0,8]\cup [24,32)$, or has two zero autocorrelation zones
of length $8$.

\end{example}

\begin{example}
Let $q=2$ and $m=5$. Let $\pi=(143)$, $c_1=0$, $c_2=0$ and
$s_i^{(1)}=1+i_{\pi(2)}+i_{\pi(3)}$. Then such $16$-QAM Golay
sequence $A$ defined in Theorem \ref{th Li} has no a ZACZ of length
$8$.

\end{example}

\begin{example}
Let $q=2$ and $m=5$. Let $\pi=(12)$, $c_1=2$, $c_2=0$ and
$s_i^{(1)}=1+i_{\pi(2)}+i_{\pi(3)}$. Then such $16$-QAM Golay
sequence $A$ defined in Theorem \ref{th Li} has no a ZACZ of length
$8$.

\end{example}

\begin{remark}
The sequences constructed in Theorems \ref{th 4} - \ref{th 6} belong
to the first two constructions in (\ref{eq qam}). For the third
construction, the above three examples show that it may have some
classes of $4^q$-QAM Golay complementary sequence with a large ZACZ.
By computer search, we found that, if $q=2$, $m=\{4, 5\}$,
$\pi(1)=1$, $\pi(2)=2$ and $2c_1=0$, and
$2d_0^{(1)}+d_1^{(1)}+d_2^{(1)}=0$, then the $16$-QAM Golay
complementary sequence $A$ defined by \eqref{eq qam} has
$R_A(\tau)=0$ for $\tau\in (0,2^{m-2}]\cup [3\cdot 2^{m-2},2^m)$,
two zero autocorrelation zones of length $2^{m-2}$. However, the
techniques that we used in Section 3 and in this section cannot
apply to this case.
\end{remark}

We summarized all the results obtained in Sections $3$ and $4$ in Table \ref{tab 9}.

\begin{table}[ht]
\begin{center}
\caption{Parameters of Golay or QAM Golay complementary sequences
with zero autocorrelation zone property}\label{tab 9} \vspace{-4mm}
\begin{lrbox}{\tablebox}
\begin{tabular}{|c|c|c|}
\hline  Permutation $\pi$  &  $(c_1,c_2)\in \Z_H\times \Z_H$ &  Zero Autocorrelation Zone \\
\hline\hline
$\pi(1)=1$,$\pi(2)=2$  &  $2c_1=0$  & $(0,2^{m-2}]$ , $ [3\cdot
2^{m-2},2^m)$ \\
\hline
$\pi(m)=1$,$\pi(m-1)=2$  &  $2c_1=0$  & $(0,2^{m-2}]$ , $ [3\cdot
2^{m-2},2^m)$ \\
\hline
$\pi(2)=2$, $\pi(3)=1$, $\pi(4)=3$  &  $2c_1=0,c_1=2c_2$  & $(0,2^{m-2}]$, $ [3\cdot
2^{m-2},2^m)$ \\
\hline
$\pi(m-1)=2$, $\pi(m-2)=1$, $\pi(m-3)=3$  &  $2c_1=0,c_1=2c_2$  & $(0,2^{m-2}]$, $ [3\cdot
2^{m-2},2^m)$ \\
\hline
$\pi(1)=2$, $\pi(2)=1$, $\pi(3)=3$  &  $2c_1=0,c_1=2c_2+t$  & $(0,2^{m-2}]$, $ [3\cdot
2^{m-2},2^m)$ \\
\hline
$\pi(m)=2$, $\pi(m-1)=1$, $\pi(m-2)=3$  &  $2c_1=0,c_1=2c_2+t$  & $(0,2^{m-2}]$, $ [3\cdot 2^{m-2},2^m)$ \\
\hline
$\pi(1)=2$, $\pi(2)=1$, $\pi(3)=3$  &  $2c_1=0,c_1=2c_2$  & $[2^{m-2},3\cdot 2^{m-2}] $\\
\hline
$\pi(m)=2$, $\pi(m-1)=1$, $\pi(m-2)=3$  &  $2c_1=0,c_1=2c_2$  & $[2^{m-2},3\cdot 2^{m-2}]$ \\
\hline
$\pi(1)=1$, $\pi(2)=3$, $\pi(3)=2$  &  $2c_1=0$  & $(0,2^{m-3}]$, $[3\cdot 2^{m-3}, 5\cdot3^{m-3}]$, $[7\cdot 2^{m-3},2^m)$ \\
\hline
$\pi(m)=1$, $\pi(m-1)=3$, $\pi(m-2)=2$  &  $2c_1=0$  & $(0,2^{m-3}]$, $[3\cdot 2^{m-3}, 5\cdot3^{m-3}]$, $[7\cdot 2^{m-3},2^m)$ \\
\hline
$\pi(1)=1$, $\pi(2)=3$, $\pi(m)=2$  &  $2c_1=0$  & $(0,2^{m-3}]$, $[3\cdot 2^{m-3}, 5\cdot3^{m-3}]$, $[7\cdot 2^{m-3},2^m)$ \\
\hline
$\pi(m)=1$, $\pi(m-1)=3$, $\pi(1)=2$  &  $2c_1=0$  & $(0,2^{m-3}]$, $[3\cdot 2^{m-3}, 5\cdot3^{m-3}]$, $[7\cdot 2^{m-3},2^m)$ \\
\hline
$\pi(1)=2$, $\pi(2)=4$, $\pi(3)=1$, $\pi(4)=3$  &  $2c_1=0,c_1=2c_2$  & $(0,2^{m-3}]$, $[3\cdot 2^{m-3}, 5\cdot3^{m-3}]$, $[7\cdot 2^{m-3},2^m)$ \\
\hline
$\pi(m)=2$, $\pi(m-1)=4$, $\pi(m-2)=1$, $\pi(m-3)=3$  & $2c_1=0,c_1=2c_2$  & $(0,2^{m-3}]$, $[3\cdot 2^{m-3}, 5\cdot3^{m-3}]$, $[7\cdot 2^{m-3},2^m)$ \\
\hline
$\pi(1)=2$, $\pi(2)=3$, $\pi(3)=1$, $\pi(4)=4$  &  $2c_1=0,c_1=2c_2$  & $(0,2^{m-3}]$, $[3\cdot 2^{m-3}, 5\cdot3^{m-3}]$, $[7\cdot 2^{m-3},2^m)$ \\
\hline
$\pi(m)=2$, $\pi(m-1)=3$, $\pi(m-2)=1$, $\pi(m-3)=4$  &  $2c_1=0,c_1=2c_2$  & $(0,2^{m-3}]$, $[3\cdot 2^{m-3}, 5\cdot3^{m-3}]$, $[7\cdot 2^{m-3},2^m)$ \\
\hline
\end{tabular}
\end{lrbox}
\scalebox{0.82}{\usebox{\tablebox}}
\end{center}
\end{table}

\section{Examples}

In the previous two sections, we have showed there exists a large
ZACZ for certain Golay sequences and QAM Golay complementary
sequences. With selected permutations $\pi$ and affine
transformations $\sum_{k=1}^mc_{\pi(k)}+c_0$, these sequences have a
large ZACZ, which can be divided into the following three cases.
\begin{enumerate}
\item  [(i)] $R_a(\tau) = 0$ and $R_A(\tau) = 0$  for $\tau\in (0,2^{m-2}]\cup [3\cdot
2^{m-2},2^m)$.
\item  [(ii)] $R_a(\tau) = 0$ and $R_A(\tau) = 0$ for
$\tau\in [2^{m-2},3\cdot 2^{m-2}]$.
\item  [(iii)] $R_a(\tau) = 0$ and $R_A(\tau) = 0$ for $\tau\in (0,2^{m-3}]\cup [3\cdot 2^{m-3}, 5\cdot
3^{m-3}]\cup [7\cdot 2^{m-3},2^m)$.
\end{enumerate}

In this section, we'll use empirical results to demonstrate these
three categories of ZACZ. A total of $6$ Golay sequences of length
$32$ labeled by $A_{1},\cdots,A_{6}$ are given in Table
\ref{ZACZ_examples}.

\begin{table}[ht]
\begin{center}
\caption{Examples of binary or quaternary Golay sequences of length
$32$ with their Autocorrelation}
\label{ZACZ_examples}
\begin{tabular}{|c|c|}
\hline  Condition  &  $\pi=(1)
$, $(c_0,c_1,c_2,c_3,c_4,c_5)=(0, 0, 1 ,1, 0, 0)$, $H=2$ \\
\hline
 Sequence  &  $A_{1}=(0, 0, 0, 1, 1, 1, 0, 1, 1, 1,
1, 0, 1, 1, 0, 1, 0, 0, 0, 1, 1, 1, 0, 1, 0, 0, 0, 1, 0, 0, 1, 0)$ \\

\hline $\{R_{A_{1}}(\tau)\}_1^{31}$ & $(0, 0, 0, 0, 0, 0, 0, 0,
-4, 0, -4, 0,-12, 0, 4, 0, 4, 0, -12, 0, -4, 0, -4, 0, 0, 0, 0, 0,
0, 0, 0)$  \\
\hline\hline

Condition  &  $\pi=(1)
$, $(c_0,c_1,c_2,c_3,c_4,c_5)=(0, 0, 1 ,1, 0, 0)$, $H=4$ \\
\hline
 Sequence  &  $A_{2}=(0, 0, 0, 2, 1, 1,3, 1, 1, 1,
1, 3, 0, 0, 2, 0, 0, 0, 0, 2, 1, 1, 3, 1, 3, 3, 3, 1, 2, 2, 0, 2)$ \\

\hline $\{R_{A_{2}}(\tau)\}_1^{31}$ & $(0, 0, 0, 0, 0, 0, 0, 0,
4j, 0, 4j, 0,12j, 0,-4j,  0, 4j, 0, -12j, 0, -4j, 0, -4j, 0, 0, 0,
0, 0, 0, 0, 0)$  \\
\hline\hline

Condition  &  $\pi=(12) $, $(c_0,c_1,c_2,c_3,c_4,c_5)=(0, 0, 0 ,0,
0,
1)$, $H=2$ \\
\hline
 Sequence  &  $A_{3}=(0, 1, 0,0,
0,1, 1,1, 0, 1, 0, 0, 0, 1, 1, 1, 0, 1, 0, 0, 1, 0, 0, 0, 1, 0, 1,
1, 0, 1, 1, 1)$ \\

\hline $\{R_{A_{3}}(\tau)\}_1^{31}$ & $(-4, 0, -4, 0, -12, 0, 4,
0, 0, 0, 0, 0, 0, 0, 0, 0, 0, 0, 0, 0, 0, 0, 0, 0,
4, 0, -12, 0, -4, 0, -4)$  \\
\hline\hline

Condition  &  $\pi=(12) $, $(c_0,c_1,c_2,c_3,c_4,c_5)=(0, 0, 0 ,0,
0,
1)$, $H=4$ \\
\hline
 Sequence  &  $A_{4}=(0, 1, 0,3,
0, 1,2, 1,  0, 1, 0, 3, 0, 1, 2, 1, 0, 1, 0, 3, 2, 3, 0, 3, 2, 3, 2,
1, 0, 1, 2, 1)$ \\

\hline $\{R_{A_{4}}(\tau)\}_1^{31}$ & $(4j, 0, 12j, 0, -4j, 0,
4j,0, 0,0, 0,  0, 0, 0, 0, 0, 0, 0, 0, 0, 0, 0, 0,
0, -4j, 0, 4j, 0, -12j,0, -4j)$  \\
\hline\hline

Condition  &  $\pi=(23) $, $(c_0,c_1,c_2,c_3,c_4,c_5)=(0, 0, 0 ,0,
0,
1)$, $H=2$ \\
\hline
 Sequence  &  $A_{5}=(0, 1, 0, 0, 0,
1, 0, 0, 0, 1, 1,  1, 1, 0, 0, 0, 0, 1, 0, 0, 1, 0, 1, 1, 0, 1, 1,
1, 0, 1, 1, 1)$ \\

\hline $\{R_{A_{5}}(\tau)\}_1^{31}$ & $(0, 0, 0, 0, -4, 0, -4,
0,-12, 0, 4,0, 0, 0,  0, 0, 0, 0, 0, 0, 4, 0, -12,
0, -4, 0, -4, 0, 0, 0, 0)$  \\
\hline\hline

Condition  &  $\pi=(23) $, $(c_0,c_1,c_2,c_3,c_4,c_5)=(0, 0, 0 ,0,
0,
1)$, $H=4$ \\
\hline
 Sequence  &  $A_{6}=(0, 1, 0, 3,
0, 1,0, 3, 0, 1, 2, 1, 2, 3, 0, 3, 0, 1, 0, 3, 2, 3, 2, 1, 0, 1, 2,
1, 0, 1, 2, 1)$ \\

\hline $\{R_{A_{6}}(\tau)\}_1^{31}$ & $(0, 0, 0, 0, 4j, 0, 12j, 0,
-4j,0, 4j, 0,0, 0, 0, 0, 0, 0, 0, 0, -4j, 0,
4j, 0, -12j,0, -4j, 0, 0, 0, 0)$  \\
\hline

\end{tabular}
\end{center}
\end{table}

\begin{figure}[ht]
\begin{minipage}[b]{0.5\linewidth}
\centering
\includegraphics[scale=0.6]{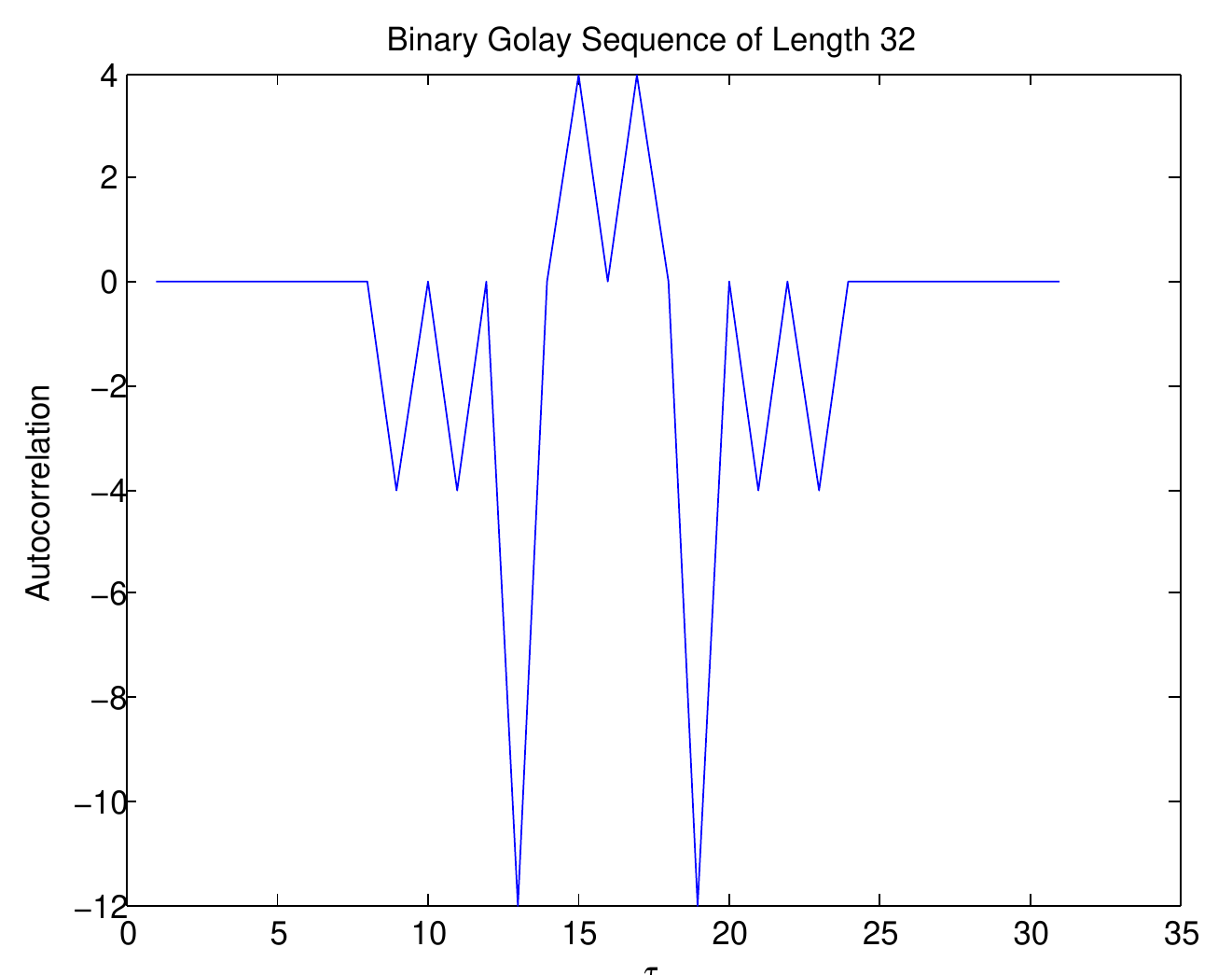}
\caption{The Autocorrelation of $A_{1}$} \label{ex1}
\end{minipage}
\hspace{0.5cm}
\begin{minipage}[b]{0.5\linewidth}
\centering
\includegraphics[scale=0.6]{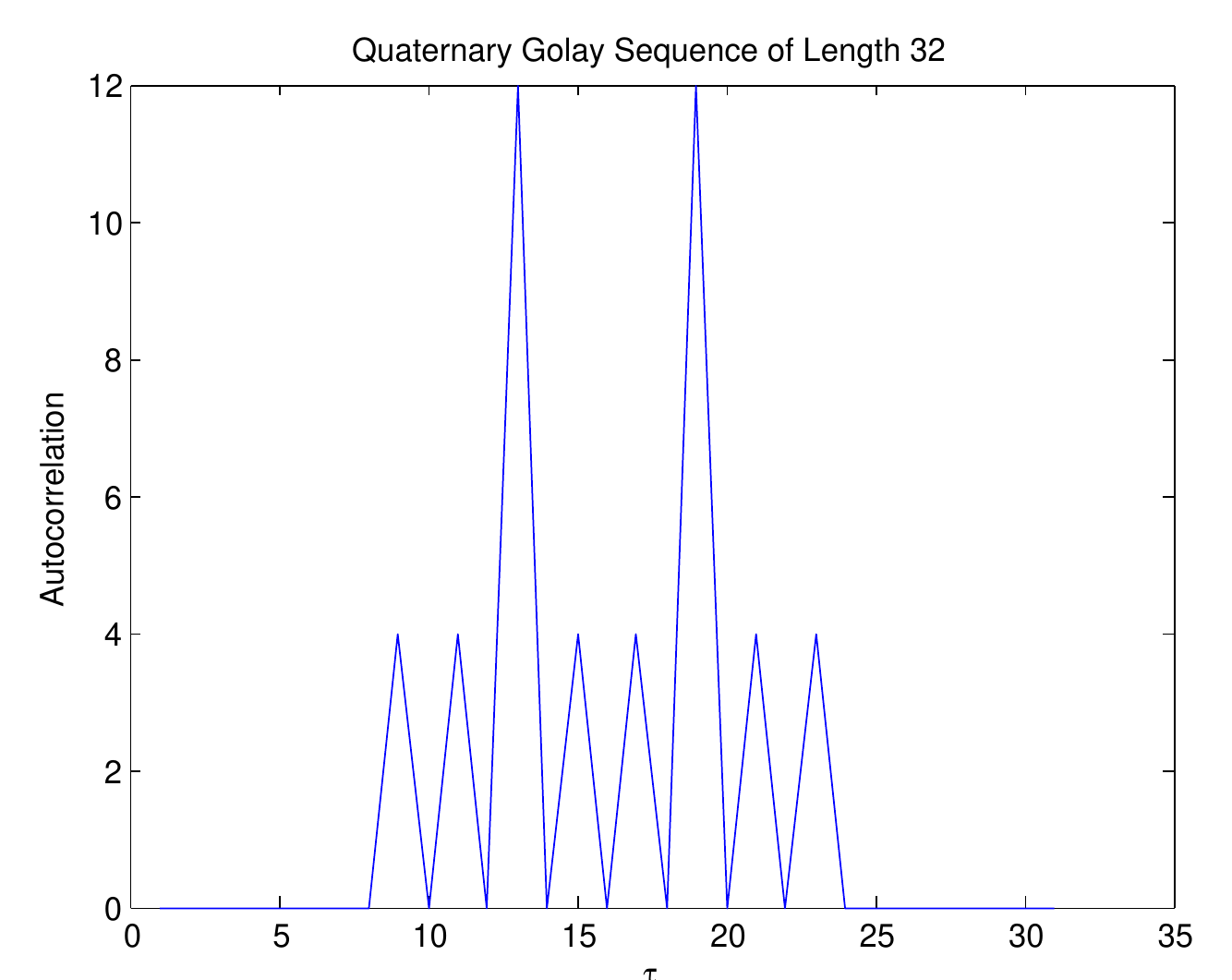}
\caption{The Autocorrelation of $A_{2}$} \label{ex2}
\end{minipage}
\end{figure}

\begin{figure}[ht]
\begin{minipage}[b]{0.5\linewidth}
\centering
\includegraphics[scale=0.6]{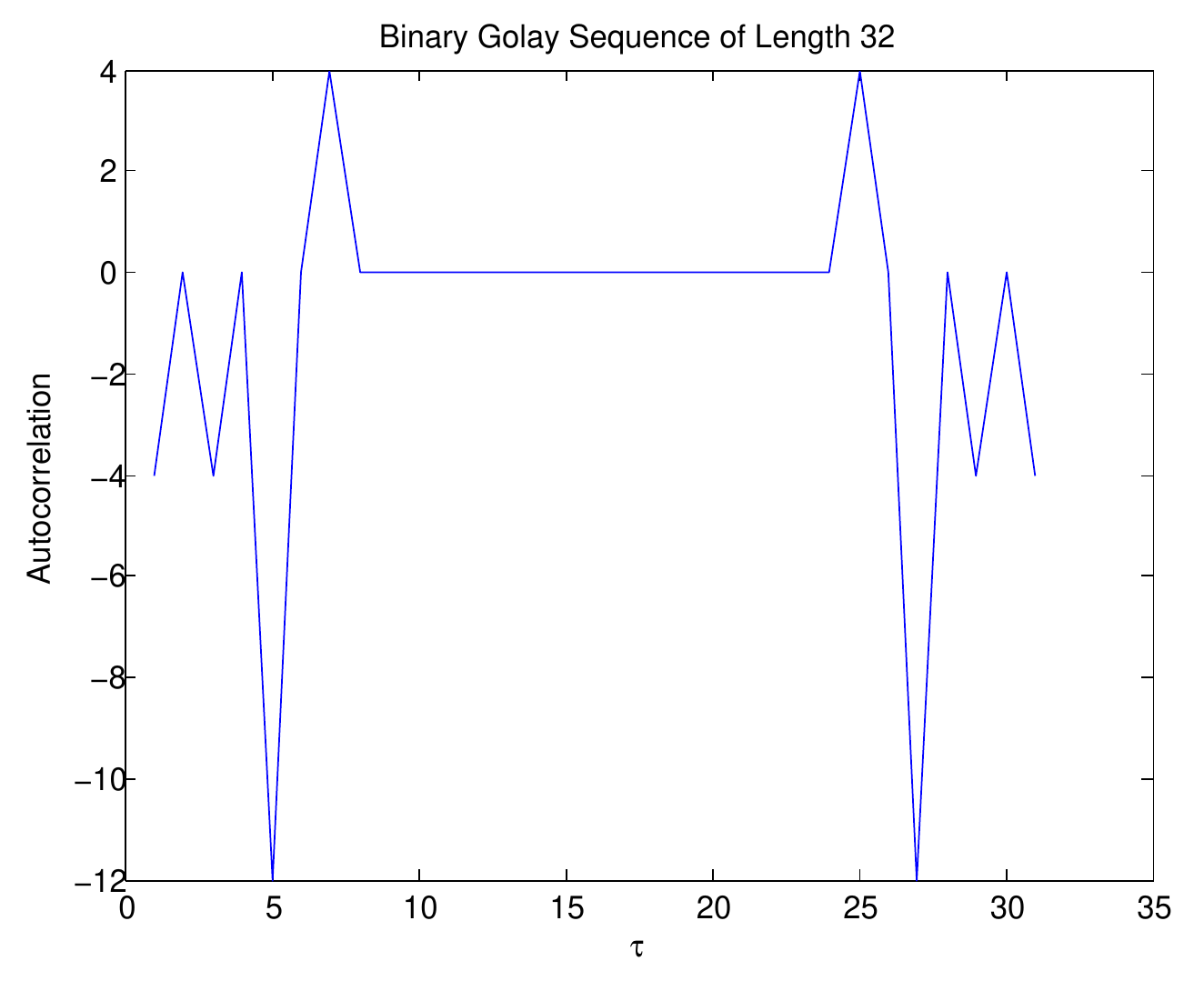}
\caption{The Autocorrelation of $A_{3}$} \label{ex5}
\end{minipage}
\hspace{0.5cm}
\begin{minipage}[b]{0.5\linewidth}
\centering
\includegraphics[scale=0.6]{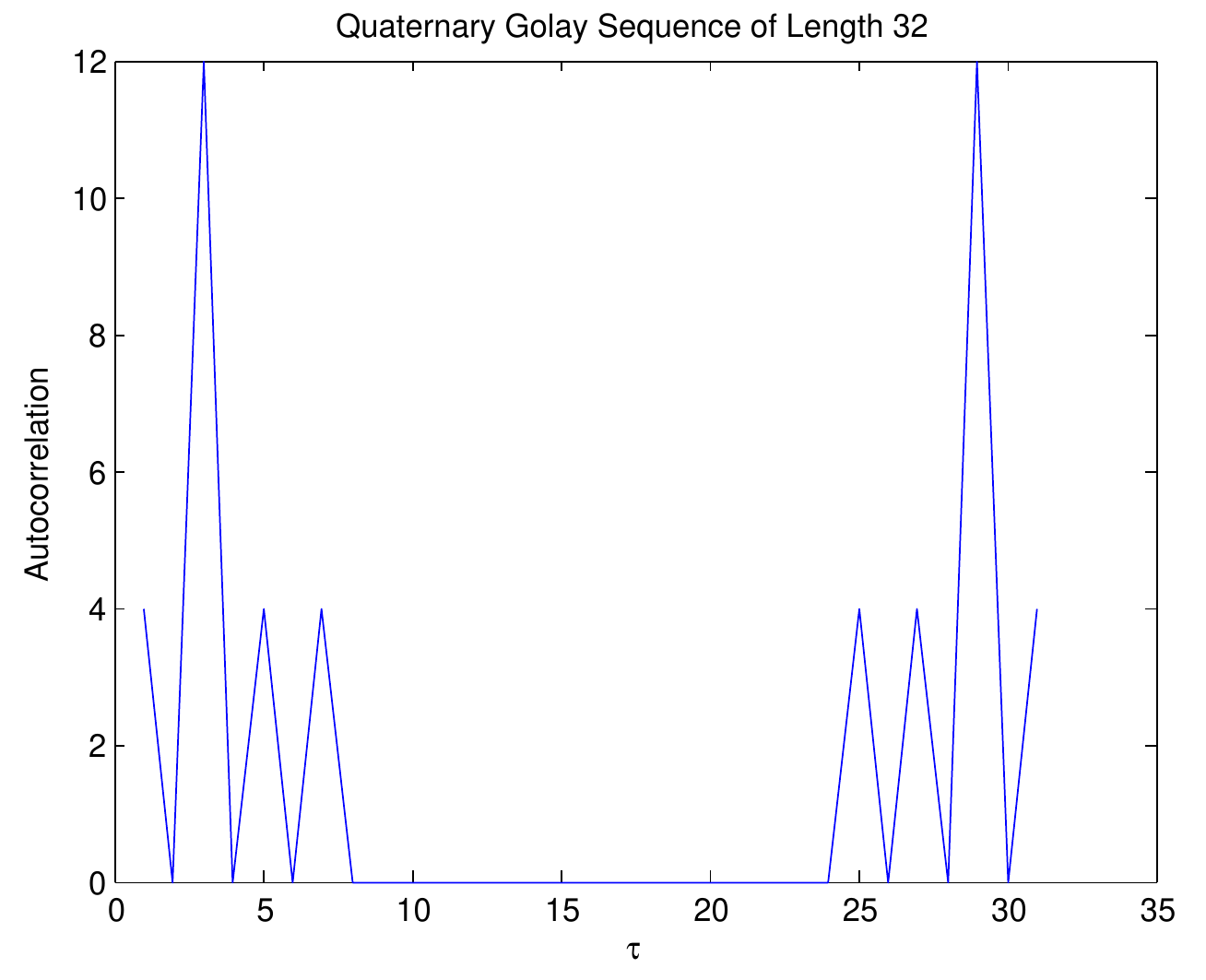}
\caption{The Autocorrelation of $A_{4}$} \label{ex6}
\end{minipage}
\end{figure}

\begin{figure}[ht]
\begin{minipage}[b]{0.5\linewidth}
\centering
\includegraphics[scale=0.6]{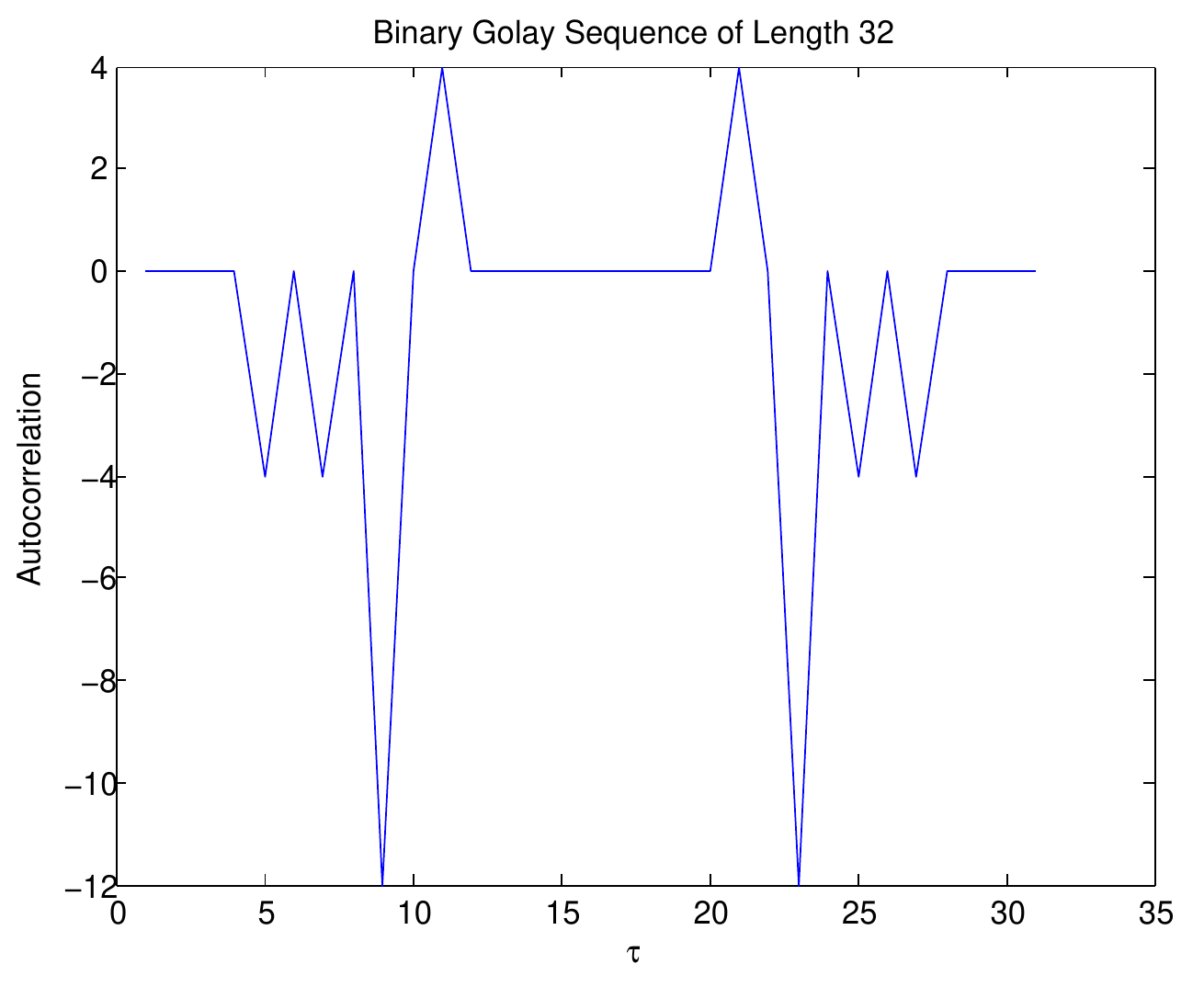}
\caption{The Autocorrelation of $A_{5}$} \label{ex3}
\end{minipage}
\hspace{0.5cm}
\begin{minipage}[b]{0.5\linewidth}
\centering
\includegraphics[scale=0.6]{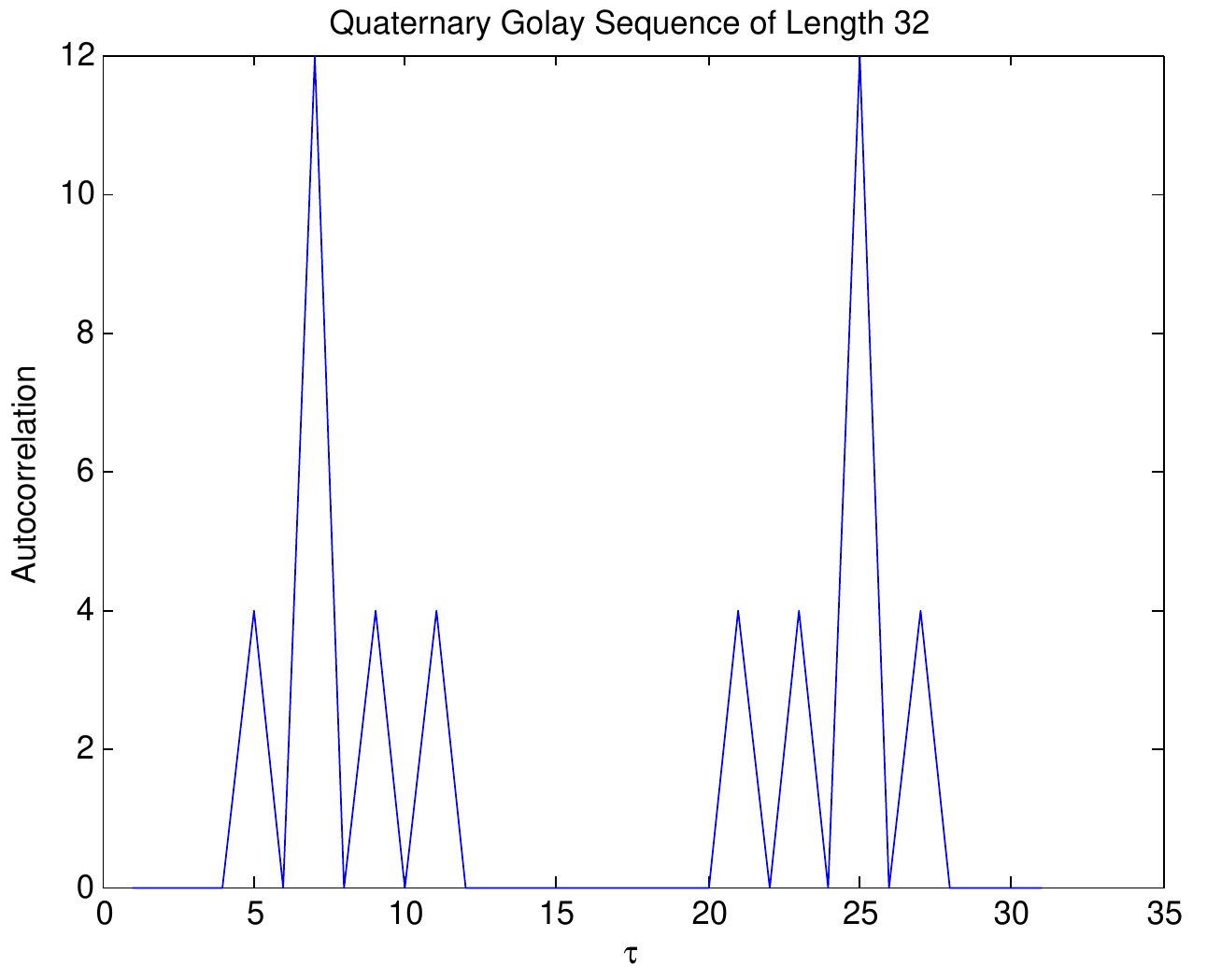}
\caption{The Autocorrelation of $A_{6}$} \label{ex4}
\end{minipage}
\end{figure}

The sequences $A_{1}$, $A_{3}$ and $A_{5}$ are binary Golay
sequences. With the same permutation and coefficients of linear
terms in \eqref{eq a}, by only changing $H$ from 2 to 4, we obtain
three quaternary Golay sequences in $A_2$, $A_4$ and $A_6$. Note
that for the figures of quaternary Golay sequences, autocorrelation
is graphed in the form of magnitude, because they contain both real
and imaginary parts. We can observe that all $6$ sequences contain a
large ZACZ. Moreover, each quaternary Golay sequence has exactly the
same ZACZ trend as its corresponding binary case. Both $A_{1}$ and
$A_{2}$ have two ZACZs of length $8$ around the two sides of the origin.
Both $A_{3}$ and $A_{4}$ have a ZACZ of length $17$ in the middle,
while $A_{5}$ and $A_{6}$ have three ZACZs: two ZACZs of length $4$
on both around the two sides of origin and one ZACZ of length $9$ in
the middle.


\section{Conclusions and Discussions}

In this paper, we have shown several constructions of GDJ Golay
sequences over $\Z_H$ and $4^q$-QAM Golay complementary sequences
which contain a large zero autocorrelation zone, where $H\ge 2$ is
an arbitrary even integer and $q\ge 2$ is an arbitrary integer.
Sequences with large ZACZ property can have wide implications in
many areas. Potential applications include system synchronization,
channel estimation and construction of signal set. This can be
briefly illustrated as follows.

{\em Synchronization}: The synchronization of the signal is
equivalent to computing its own autocorrelations \cite{Haykin,
Pursley}. If the signal delay does not exceed of the ZACZ, then
early synchronization or late synchronization will introduce no
interference to the system. There will only be a peak value at the
origin (i.e., correct synchronization). Thus the synchronization of
system can be achieved.

{\em Channel Estimation} : Golay sequences with large ZACZ property
can be used as pilot signals for channel estimation purposes in an
LTI system. The relationship between input $x(t)$, channel impulse
repones $h(t)$ and received signal $y(t)$ and white Gaussian noise
$n(t)$ is given by \cite{Haykin}
\begin{eqnarray}\label{LTI} y(t) = x(t) \otimes h(t) + n(t) \end{eqnarray}
where $\otimes$ is the convolution operator. Once synchronization of
signal is achieved as explained above using its large ZACZ property,
then the received signal $y(t)$ can be accurately recovered. Note
from (\ref{LTI}), we have the approximated channel impulse response
is:
\begin{eqnarray*}
Y(f)=X(f)H(f) + N(f)\,\,\Longrightarrow\,\,
\frac{Y(f)}{X(f)}=H(f)+\frac{N(f)}{X(f)}
\end{eqnarray*}
where $X(f)$, $Y(f)$ and $N(f)$ are the Fourier transforms of
$x(t)$, $y(t)$ and $n(t)$ respectively. Therefore, the approximated
channel response $\hat{h}(t)$ is:
$$\hat{h}(t)\approx \mathcal{F}^{-1}\frac{Y(f)}{X(f)}  $$
where $\mathcal{F}^{-1}$ is the inverse Fourier transform operator.

 Another possible application of Golay
sequences with large ZACZ is that it can be used to construct
spreading sequence sets for CDMA systems. This will be a future
research work.

\section*{Acknowledgment}
The work is supported by NSERC Discovery Grant.


\begin{thebibliography}{99}






\bibitem{CLH}
C. Y. Chang, Y. Li, and J. Hirata, ``New 64-QAM Golay complementary
sequences,'' {\em IEEE Trans. Inform. Theory}, vol. 56, no. 5, pp.
2479-2485, May. 2010.

\bibitem{CVT}
C. V. Chong, R. Venkataramani, and V. Tarokh, ``A new construction
of 16-QAM Golay complementary sequences,'' {\em IEEE Trans. Inform.
Theory}, vol. 49, no. 11, pp. 2953-2959, Nov. 2003.

\bibitem{DJ99}
J.A. Davis and J. Jedwab, ``Peak-to-mean power control in OFDM,
Golay complementary sequences and Reed-Muller codes,'' {\em IEEE
Trans. Inform. Theory}, vol. 45, no. 7, pp. 2397-2417, Nov. 1999.

\bibitem{DF00}
X. M. Deng and P. Z. Fan, ``Spreading sequence sets with zero
correlation zone,'' {\em Electron. Lett.}, vol. 36, no. 11 pp.
993-994, May 2000.



\bibitem{FH00} P. Z. Fan and L. Hao, ``Generalized orthogonal sequences and
their applications in synchronous CDMA system,'' {\em IEICE Trans.
Fundam.}, vol. E83-A, no. 11, pp. 1-16, Nov. 2000.

\bibitem{Golay}
M. J. E. Golay, ``Complementary series,''  {\em IRE Trans. Inform.
Theory, vol. IT-7}, no. 2, pp. 82-87, Apr. 1961.

\bibitem{GG05}
S. W. Golomb and G. Gong, {\em Signal Designs With Good Correlation:
For Wireless Communication, Cryptography and Radar Applications.}
Cambridge, U.K: Cambridge Univeristy Press, 2005.

\bibitem{Gong}
G. Gong, F. Huo, and Y. Yang, ``Large zero autocorrelation  zone of
Golay sequences,'' {\em IEEE Globe Communications Conference 2011},
Houston, Texas, USA, Dec. 5-9th, 2011, submitted.

\bibitem{Hayashi02}
T. Hayashi, ``Binay sequences with orthogonal subsequences and a
zero-correlation zone: Pair-preserving shuffled sequences,'' {\em
IEICE Trans. Fundam.}, vol. E85-A, no. 6, pp. 1420-1425, 2002.

\bibitem{Hayashi04}
T. Hayashi, ``A generalization of binary zero-correlation zone
sequence sets constructed from Hadamard matrices,'' {\em IEICE
Trans. Fundam.}, vol. E87-A, no. 1, pp. 559-565, 2004.

\bibitem{Haykin}
S. Haykin and M. Moher.
\newblock {\em Communication Systems}.
\newblock John Wiley \& Sons, U.S, 2009.

\bibitem{HG10}
H.G. Hu and G. Gong, ``New sets of zero or low correlation zone
sequences via interleaving techniques,'' {\em IEEE Trans. Inform.
Theory}, vol. 56, no. 4, pp. 1702-1713, April 2010.


\bibitem{Lee}
H. Lee and S.W. Golomb, ``A new construction of 64-QAM Golay
complementary sequences,'' {\em IEEE Trans. Inform. Theory}, vol.
52, no. 4, pp. 1663-1670, April 2006.

\bibitem{Li2008}
Y. Li, ``Commnents on ``A new construction of 16-QAM Golay
complementary sequences'' and extension for 64-QAM Golay
sequences,'' {\em IEEE Trans. Inform. Theory}, vol. 54, no. 7, pp.
3246-3251, July 2008.

\bibitem{Li2010} Y. Li, ``A construction of general QAM Golay complementary sequences,''
{\em IEEE Trans. Inform. Theory}, vol. 56, no. 11, pp. 5765-5771,
May 2010.

\bibitem{Long}
B. Long, P. Zhang, and J. Hu, ``A generalized QS-CDMA system and the
design of new spreading codes,'' {\em IEEE Trans. Veh. Tech.}, vol.
47, pp. 1268-1275, 1998.



\bibitem{RC06} A. Rathinakumar and A. K. Chaturvedi, ``A new framework for
constructing mutually orthogonal complementary sets and ZCZ
sequences,'' {\em IEEE. Trans. Inform. Theory}, vol. 52, no. 8, pp.
3817-3826, Aug. 2006.

\bibitem{RC08} A. Rathinakumar and A. K. Chaturvedi, ``Complete mutually
orthogonal Golay complementary sets from Reed-Muller codes,'' {\em
IEEE. Trans. Inform. Theory}, vol. 54, no. 3, pp. 1339-1346, Mar.
2008.

\bibitem{Paterson}
K.G. Paterson, ``Generalized Reed-Muller codes and power control for
OFDM modulation,'' {\em IEEE. Trans. Inform. Theory}, vol. 46, no.
1, pp. 104-120, Feb. 2000.

\bibitem{Pursley}
M.B. Pursley.
\newblock {\em A Introduction to Digital Communications}.
\newblock Pearson Prentice Hall,  U.S, 2005.


\bibitem{seberry}
J. R. Seberry, B. J. Wysocki, and T. A. Wysocki, ``On a use of Golay
sequences for asynchronous DS CDMA applications,'' {\em Advanced
Signal Processing for Communication Systems The International Series
in Engineering and Computer Science},  Vol. 703, pp. 183-196, 2002.



\bibitem{Tang00}
X. H. Tang, P. Z. Fan, and S. Matsufuji, ``Lower bounds on the
maximum correlation of sequence set with low or zero correlation
zone,'' {\em Electron. Lett.}, vol. 36, pp. 551-552, Mar. 2000.

\bibitem{Tang06}
 X. H. Tang and W. H. Mow, ``Design of spreading codes for
quasisynchronous CDMA with intercell interference,'' {\em IEEE J.
Sel. Areas Commun.}, vol. 24, no. 1, pp. 84-93, Jan. 2006.



\bibitem{van Nee}
R. D. J. van Nee, ``OFDM codes for peak-to-average power reduction
anderror correction,'' {\em in Proc. IEEE GLOBECOM}, London, U.K,
pp. 740-744, Nov. 1996.

\bibitem{Wilkinson95}
T. A. Wilkinson and A. E. Jones, ``Minimization of the peak to mean
envelope power ratio of multicarrier transmission schemes by block
coding,'' {\em in Proc. IEEE 45th Vehicular Technology Conf.},
Chicago, IL,  pp. 825-829, Jul. 1995.



\bibitem{Wilkinson96}
T. A. Wilkinson and A. E. Jones, ``Combined coding for error control
and increasedrob ustness to system nonlinearities in OFDM,'' {\em in
Proc. IEEE 46th Vehicular Technology Conf.}, Atlanta, GA, pp.
904-908, 1996.

\bibitem{Wulich}
D.Wulich, ``Reduction of peak to mean ratio of multicarrier
modulation using cyclic coding,'' {\em Electron. Lett.}, vol. 32,
pp. 432-433, 1996.



\bibitem{ZTG10} Z.C. Zhou, X.H. Tang, and G. Gong, ``A new class of sequences
with zero or low correlation zone based on interleaving technique,''
{\em IEEE Trans. Inform. Theory}, vol. 54, no. 9, pp. 4267-4273,
April 2008.

\end{thebibliography}
\end{document}